\documentclass[a4paper,11pt]{article}

\usepackage{amssymb}
\pdfoutput=1 

\usepackage{jcappub} 
\usepackage[T1]{fontenc} 
\usepackage{adjustbox}
\usepackage{float}
\usepackage{xcolor}
\usepackage{hyperref}
\usepackage{enumitem}
\usepackage{graphicx} 
\usepackage{makecell} 
\usepackage{caption} 
\usepackage{textcomp}
\usepackage{tikz}
\usepackage{booktabs} 
\usepackage[T1]{fontenc}
\usepackage[utf8]{inputenc} 
\usepackage{amssymb}
\usepackage{siunitx} 
\usepackage{amsmath} 
\usepackage{amsmath}  
\usepackage{mathrsfs} 
\usepackage{amssymb}  
\usepackage{subcaption}
\usepackage{empheq}
\usepackage{cleveref}

\newcommand{\be}{\begin{equation}}
\newcommand{\ee}{\end{equation}}
\newcommand{\ba}{\begin{eqnarray}}
\newcommand{\ea}{\end{eqnarray}}

\title{
Probing Dark Matter-Electron Interactions in the Cosmic Microwave Background Radiation}

\author[a,1]{Rahul Dhyani,}
\author[b,1]{Arnab Paul,}
\author[a,2]{Arindam Chatterjee}

\affiliation[a]{Department of Physics, Shiv Nadar Institution of Eminence Deemed to be University, Gautam Budhha 
Nagar, Uttar Pradesh 201314, India}
\affiliation[b]{Centre for Strings, Gravitation and Cosmology, Department of Physics, Indian Institute of Technology Madras, Chennai 600036, India}
\emailAdd{rahul.physics123@gmail.com}
\emailAdd{arnabpaul9292@gmail.com}
\emailAdd{arindam.chatterjee@snu.edu.in}

\abstract{In this article, we consider Dark Matter (DM) interactions and study the same in 
the light of the Cosmic Microwave Background Radiation (CMBR) data. In particular, we 
focus on the DM-electron interactions. Assuming that such interactions are mediated by 
rather heavy mediators, we consider effective operators describing the relevant interaction 
terms in the lagrangian. The presence of such interaction terms leads to both DM 
annihilation and DM-electron scattering (drag). We focus on operators which lead to 
velocity-independent DM annihilation and DM-electron scattering cross-sections.  Using the 
CMBR data, we study the implications of both of these effects, imposing constraints on the 
respective effective operators. This analysis underscores the importance of taking both 
scattering and annihilation processes into consideration in the study of DM interactions. We 
observe that the constraints on the DM annihilation and scattering cross-sections can 
change, up to about 13\% and 12\%, respectively, for the benchmark scenarios we 
considered, depending on the mass of DM, as compared to the scenario where only DM 
annihilation is accounted for.}

\begin{document}

\maketitle
\tableofcontents

\newpage
 
\section{Introduction} \label{sec:Introduction}
The standard model of cosmology, the $\Lambda$CDM model, has been successful in 
light of various observational data, including the anisotropies of the Cosmic Microwave 
Background (CMBR) \cite{Planck:2015fie, Planck:2018vyg}. In the simplest version, 
this model incorporates the cosmological constant $\Lambda$  and Cold Dark Matter
(CDM) species, which does not interact with the constituent particles of the Standard 
Model (SM) of particle physics. It is estimated that CDM and $\Lambda$ account for 
approximately 26\% and 69\%  \cite{Planck:2018vyg} of the current energy budget of 
our Universe, respectively.

While there is no suitable candidate for DM within the SM of particle physics, several 
well-motivated extensions of the SM incorporate particle DM candidates; see, e.g. 
refs. \cite{D1, Jungman:1995df,Cirelli:2024ssz} for reviews. In most theoretical frameworks, 
DM possesses non-vanishing interactions with the SM particles. For instance, Weakly 
Interacting Massive Particles (WIMP), which have been well studied in the literature 
as DM candidates, possess sizable interactions with the SM particles. Adequate 
interaction with the SM particles is essential for the thermal production of such 
DM candidates in the very early Universe \cite{Kolb:1990vq}. 

The possibility of interactions between DM and SM particles has been widely 
explored in the search for DM. Both direct detection experiments 
\cite{PICO:2019vsc,Emken:2019tni,PandaX-II:2020oim,LZ:2022lsv,XENON:2023cxc} 
and indirect searches \cite{IceCube:2021xzo, Fermi-LAT:2016uux,Fermi-LAT:2015att,MAGIC:2016xys,Regis:2021glv,Adriani:2012paa,Adriani:2008zq,AMS:2016oqu,Caroff:2016abl,PhysRevLett.113.221102,Vecchi:2017uzt,Cirelli:2023tnx,Cirelli:2024ssz} of DM constrain such interactions. Further, within specific theoretical 
frameworks, signatures of DM in the form of missing energy and momentum have 
been probed at high-energy colliders \cite{CMS:2024zqs,CMS:2023bay,ATLAS:2024rlu,ATLAS:2024npu,ATLAS:2021kxv,ATLAS:2020uiq,ATLAS:2022tjh}, which constrains the  
production cross-section of DM particles, and thus, the interaction strengths of SM and 
DM particles. It is important to note that the search for particle DM candidates in direct detection 
experiments can generally be challenging for rather light (sub-GeV) DM candidates, 
due to the rather small momentum transfer to the nuclei. In this context, the interaction of 
DM with electrons in the heavy atoms (see e.g. \cite{Essig:2012yx, Essig:2017kqs}), 
or in the semiconductor materials (e.g. in silicon \cite{SENSEI:2019ibb, SENSEI:2020dpa}) 
and/or with various collective excitations have been explored in different materials, 
see e.g. ref.~\cite{Kahn:2021ttr} for review. In particular, stringent constraints on 
DM-electron interaction for rather light DM from the direct searches 
\cite{DarkSide:2018ppu,SENSEI:2019ibb, SENSEI:2020dpa} have been obtained.\footnote{Note that, for 
DM to reach the detectors, its interaction with SM particles should be sufficiently 
weak so that it traverses through Earth's atmosphere and the experimental shielding 
\cite{Zaharijas:2004jv,Emken:2017qmp,Mahdawi:2017cxz,Hasenbalg:1997hs,Emken:2019tni,c1,c2}.}

In the era of precision cosmology, cosmological data have been used in the literature 
to probe and constrain DM interactions. In particular, the effect of DM annihilation 
\cite{Padmanabhan:2005es,slatyer09,Galli09,Lopez-Honorez:2013cua,Dvorkin13,lesgour13,Slatyer:2015jla,Green:2018pmd,Bertschinger:2006nq,Kawasaki:2021etm,Wang:2025tdx}
and scattering with the baryonic matter (and electrons) \cite{Dvorkin13,Xu:2018efh,Gluscevic:2017ywp,kim18,Boddy:2018wzy,dvorkin20,kim21,Boddy:2022tyt,Ali-Haimoud:2023pbi} on the 
CMBR anisotropies have been well-studied in the literature. To be specific, DM annihilation 
into visible matter during and around the recombination epoch can lead to ionization 
and excitation of the neutral hydrogen atoms, and heating of the baryonic plasma. This 
can enhance the duration of the recombination epoch.  Further, DM annihilation can lead 
to an enhanced electron-positron number density around the same epoch, which affects 
the polarization of CMBR photons. Apart from this, DM annihilation can leave its imprint on 
the CMBR spectral distortion as well \cite{Ali-Haimoud:2015pwa,Chluba:2018cww,Ali-Haimoud:2021lka}. 
The scattering of SM particles, in particular, protons, neutrons (which are in the helium atoms), and electrons, around the recombination epoch also leaves its imprint in the CMBR, as such processes affect the evolution of perturbations. The scattering process leads to cooling of the baryon temperature, and it washes out power in the small length scales. This, in turn, 
has been used to constrain such scattering or ``drag'' processes. Apart from interactions 
with the visible particles, interactions between DM  with neutrinos 
\cite{N1, N2, N3,Paul:2021ewd,Ghosh:2019tab}, Dark Radiation (DR) 
\cite{Ackerman:2008kmp} and Dark Energy (DE) 
\cite{Asghari:2019qld,Lucca:2021dxo,BeltranJimenez:2021wbq,Poulin:2022sgp} 
have been studied in different contexts in the light of CMBR data. Note that, during the recombination epoch, the evolution of density, as well as the peculiar velocities are governed by (linear) perturbation theory, and thus, computation of the power spectrum involves linear calculations. This enables the detection of rather small deviations from the standard picture and is generally insensitive to uncertainties on the DM density 
(and velocity) Profile, unlike direct/indirect searches. It is worth mentioning that the effect of DM annihilation has been studied in the context of reionization \cite{Mapelli:2006ej,Cirelli:2009bb,valdes2010particle,Kaurov:2015lby,Slatyer:2018aqg,Liu:2016cnk,Slatyer:2015kla,Munoz:2015bca}, and proposals have been made to 
probe the same in 21 cm cosmology \cite{Valdes:2007cu}. Further, there have been 
studies on constraining annihilation of light DM from the Big Bang Nucleosynthesis (BBN) 
\cite{kolb&turner,Jedamzik:2004ip,Kawasaki:2015yya,Depta:2019lbe,Braat:2024khe}.

From the perspective of a theoretical model of particle DM, interactions with the SM particles 
are described by an interaction term in the lagrangian. Such interactions, if present, generally 
lead to both DM annihilation into a pair of SM particle and anti-particle, when the process is 
kinematically allowed; as well as the scattering of DM with the respective SM species. 
Thus, in such scenarios, it is essential to take both DM annihilation and scattering 
into consideration. In the present work, we consider DM interactions with electrons 
described by effective operators\footnote{Generally, DM particles may interact with any 
SM particles, thus, effectively interacting with protons and electrons, which are present 
around the recombination epoch. However, for example, in the leptophilic models of 
DM, it interacts with only leptons, see e.g. \cite{Gu:2017}.}. This is appropriate when the 
interactions are mediated by a rather heavy mediator, the mass of which is much 
greater than twice the mass of the DM particle. In general, there may be different 
operators at play, and different operators may dominate the annihilation and the 
scattering processes with electrons. Further, it may also happen, 
depending on the interaction terms present and the mass of the DM particle, that annihilation 
and scattering with electrons, as well as scattering with protons (and neutrons) are present. 
From the perspective of the cosmological framework, thus, the relevant extension of 
$\Lambda$CDM includes two parameters describing the effect of the DM scattering 
or drag, as well as the DM annihilation. It is important to study such a scenario, 
as it illustrates the properties of DM and provides complementary constraints on the 
same at a very different epoch (around the recombination) in the evolution of the 
Universe;  ii) it is useful to establish the robustness of the cosmological parameters 
within the base $\Lambda$CDM model and study possible degeneracies with the 
additional parameters\footnote{In the later context, it is worth noting that DM-SM particle 
scattering, if present, can relax the discrepancy \cite{DiValentino:2017oaw,He:2023dbn} 
in the determination of the $\sigma_8$ parameter, which arises while determining 
this parameter using CMBR data, as well as data from various galaxy surveys 
\cite{KiDS:2020suj,Hildebrandt:2020rno,Philcox:2021kcw,Chen:2022jzq,Zhai:2022yyk,DES:2021wwk}.}. In the present context, we study the effect of DM scattering or drag and 
annihilation into electrons, for scalar and fermionic DM candidates using a modified version of \texttt{CLASS} 
\cite{lesgourgues2011cosmic,Blas_2011}. We 
perform a Bayesian analysis using the publicly available package \texttt{MontePython} 
\cite{Brinckmann:2018cvx} to determine the upper limits on the relevant 
parameters using the CMBR data from \texttt{Planck} \cite{Planck:2018vyg}. Using 
this upper limit, we constrain the effective interaction terms in the lagrangian for 
different masses of DM particles within the range of 1 MeV - 10 GeV. For simplicity, we 
have considered the presence of one interaction term at a time while deriving the 
constraints on the relevant lagrangian parameters.

The paper is organized as follows. In Sec .~\ref{sec:pert}, we briefly discuss the effect of DM annihilation into a pair of electron-positron and the effect of DM 
scattering with electrons in the light of CMBR.In this discussion, we considered the individual effects of 
annihilation and scattering. Further, we also review the relevant 
perturbation equations in this context. In Sec .~\ref{sec:pert2}, we sketch the 
effective interaction terms in the lagrangian describing the DM-electron interactions in the context of a scalar and a fermionic DM. Following this, in Sec.~\ref{sec:result} 
the effects of DM annihilation and scattering with electrons have been discussed on the CMBR temperature and the polarization power spectra. In the 
same section, we then present our results: the constraints on the annihilation and 
drag parameters from the CMBR data, and the respective constraints on the 
DM-electron interaction terms in the lagrangian. We have used the Planck 2018 data 
set (Planck high-l TTTEEE lite, low-l TT, low-l EE, lensing). Finally, in 
Sec.~\ref{sec:conclusion} we conclude. 
 
\section{DM Interactions with the SM: Implications in the light of CMBR}
\label{sec:pert}
In this section, we review the implications of DM ($\chi$) annihilation in the early Universe
in light of CMBR. In particular, we focus on the effect of DM annihilation around the last 
scattering surface $(z \simeq 1100)$. In the presence of the interaction terms between the DM 
and the SM fields in the lagrangian, two DM particles can annihilate into SM particles, when such a process is kinematically viable. If DM particles thermalized with the SM particles in the very early Universe, as is the case for thermally produced DM particles, the DM annihilation rate becomes much smaller in comparison with the expansion rate after the thermal freeze-out. The success of the standard Big Bang  Nucleosynthesis (BBN), however, constrains the mass of DM $(m_{\chi}) \gtrsim 10$ MeV \cite{Depta:2019lbe} for DM candidates that thermalize with the SM thermal bath. In the following discussion, we will be agnostic about the production mechanism for DM. Further, note that, for DM annihilation into a pair of electron-positron pair to be kinematically viable, the minimum mass of such a DM particle needs to be greater than about 0.5 MeV, i.e. $m_{\chi} \gtrsim m_e$. We will consider $m_{\chi}$ in the range of $\mathcal{O}(1)$ MeV - $\mathcal{O}(10)$ GeV in the present context. In the following, we begin by reviewing the implications of DM annihilation into a pair of electrons and positrons in the early Universe around the epoch of recombination. 

\subsection{DM Annihilation into $\bar{e} e$}
\label{sec:dmann}
Around the epoch of recombination, DM annihilation into the SM particles, in particular into a pair of electron and positron, injects energy into the SM thermal bath. The thermal bath around this epoch consists of photons and mostly electrons, protons, and helium 
nuclei. The efficiency and redshift dependence of the energy transfer process from the DM to the SM plasma, for a specific mass of DM, generally depend on the DM annihilation channels, which we assume to be an electron and positron pair in this 
context. The energetic electrons transfer energy to the photons via inverse Compton scattering, which remains effective around the epoch of recombination. At energies $\lesssim \mathcal{O}(100)$ MeV, these produce photons, which effectively 
ionize hydrogen \cite{Padmanabhan:2005es,Ripamonti:2006gq,slatyer09,Galli09,Dvorkin13, chen,  valdes2010particle, Hutsi:2011vx}. The electrons and photons with lower 
energies participate in collisional heating, excitations, and ionization 
\cite{Padmanabhan:2005es,Galli09, Galli:2013dna}. Further, the photons with higher energies can 
initiate an electromagnetic cascade by Compton scattering with an electron, 
or escape without heating the plasma \cite{Padmanabhan:2005es} . Note that around the epoch of recombination 
($z \simeq 1100$), The inverse Compton scattering is very efficient in transferring 
the energy from relativistic electrons to the photons, as the respective rate 
is much larger than the expansion rate of the Universe \cite{Padmanabhan:2005es}. 

Quantitatively, the energy injection (per unit volume per unit time) from the annihilation 
of two DM particles, at a redshift $z$ around the recombination epoch is given by \cite{Galli09,lesgour13}, 
\begin{equation}
\frac{dE}{dV dt}(z) =  \rho_{\rm crit}^2 c^2 \Omega_{\rm c}^2 (1 + z)^6 
 P_{\mathrm{ann}} 
 \label{eq:dmann}
\end{equation}
In the above equation, $P_{\mathrm{ann}} =  f_{\rm eff}(z) \frac{\langle \sigma v \rangle} 
{m_{\chi}}$, where $ \langle \sigma v \rangle$ denotes the annihilation cross-section 
multiplied by the relative speed of the DM particles averaged using the phase space distribution function of DM particles at redshift $z$, $f_{\rm eff}(z)$ denotes the effective fraction of the injected energy absorbed by the SM plasma at redshift $z$ \footnote{The efficiency of energy injection process in the SM plasma, as a function 
of the redshift $z$, is included in the parameter $\textit{f}_\text{eff}$. As demonstrated 
in refs. \cite{Ripamonti:2006gq, Lopez-Honorez:2013cua, Slatyer:2015jla}, around redshift of $z \simeq 1100$  $f_{\rm eff}(z)$ is independent of the redshift $z$ to a very good approximation for a particular annihilation channel. A large value of $f_{\rm eff}$ corresponds to an instantaneous transfer of energy to the SM plasma, 
referred to as the ``on-the-spot" approximation. A detailed computation of the function $f(z)$, beyond the on-the-spot approximation, can be found in 
ref.~\cite{Slatyer:2015jla}.}, $\rho_{\rm crit}$ denotes the critical energy density in our Universe in the present epoch, $\Omega_{\rm c,0}$ denotes the fractional contribution to the energy density from DM at the present epoch and $c$ denotes the speed of light in vacuum.  Around the recombination epoch ($z \simeq 1100$), which is of interest in the present context, the injected energy is distributed 
approximately equally, contributing towards ionization and excitation of the neutral atoms and heating the plasma. In a fully ionized plasma, the injected energy contributes to heating the plasma. Following ref.~\cite{Chen:2003gz}, an approximate estimation of the fraction of injected energy contributing towards ionization and excitation each can be given by $(1- x_e)/3$ while $(1+2x_e)/3$ contributed towards heating of the plasma, where $x_e$ is free electron fraction. More accurate estimations for the relevant fractions have been estimated in refs. \cite{1985ApJ...298..268S,Valdes:2008cr,Evoli:2012zz,Galli:2013dna}. The 
equation governing the evolution of the temperature of the baryons and 
electrons $T_b$ is given by the Compton evolution equation, 
\begin{eqnarray}
\begin{aligned}
(1+z) \frac{d T_b}{d z}= & \frac{8 \sigma_T a_R T_{r}^4}{3 m_e c H(z)} \frac{x_e}{1+f_{H e}+x_e}\left(T_b-T_{r}\right) +2 T_b  \\
& -\frac{2}{3 k_B H(z)} \frac{\kappa_h(z)}{1+f_{He}+x_e},
\end{aligned}
\end{eqnarray}
where the term involving $\kappa_h$ corresponds to the effect of heating from 
additional energy injection in the SM plasma, and is given by, 
\begin{equation}
\kappa_h(z) = \frac{(1+ 2 x_e)}{3 n_H(z)} \frac{dE}{dV dt}(z).
\end{equation}
In the above equation, $T_{r}$ denotes the radiation (photon) temperature at redshift 
$z$, $\sigma_T$ denotes the Thomson scattering cross-section, $a_R$ is the radiation 
constant, $k_B$ is the Boltzmann constant, $m_e$ is the mass of electron, $c$ 
denotes the speed of light in vacuum, $z$ denotes the redshift. Further, $H(z)$ stands 
for the Hubble parameter at redshift $z$, $x_e$ and $f_{He}$ denote the free electron 
fraction and the fraction of Helium nuclei with respect to the number of $H$ nuclei 
($n_H(z)$) present in the SM plasma at redshift $z$. Note that at the onset of 
the recombination epoch, Thompson scattering is effective and $T_{r} \simeq T_b$. 
During the recombination epoch, the free electrons are captured by the hydrogen 
(and Helium) nuclei to form neutral atoms, and around the last scattering surface, 
the medium becomes transparent for photons, which we observe as the CMBR. 
The effect of the additional energy injection on the formation of the neutral atoms, 
and the evolution of the free electron fraction can be described to a good 
approximation by the following equation \cite{Galli09,slatyer09,lesgour13}: 
\begin{equation}
\frac{d x_e}{d z} = \frac{1}{(1+z) H(z)} \left[ R_s(z) - I_s(z) - I_X(z) \right],
\label{eq:xe}
\end{equation}
where $R_s(z)$ is the standard recombination rate, $I_s(z)$ is the ionization rate 
due to standard sources  and $I_X(z)$ is the ionization rate due to annihilating DM 
particles at redshift $z$. Considering the effective three-level-atom framework \cite{Peebles:1968ja}, to a good approximation, these are given by \cite{Galli09,slatyer09,lesgour13,Slatyer:2015jla}, 
\begin{equation}\label{eq:2.2}
\left[ R_s(z) - I_s(z) \right] = C \left[ x_e^2 n_H \alpha_B - \beta_B \left(1 - x_e \right) 
e^{-E_{21} / k_B T_r} \right],
\end{equation}
In the above equation, $n_H$ denotes the number density of hydrogen nuclei, $\alpha_B$ 
and $\beta_B$ are the effective recombination and photoionization rates from the first 
excited state  in Case $B$ recombination, $E_{21}$ stands for the energy difference 
between the $2 s$ level and the ground state of the Hydrogen atom, and $T_r$ is the 
radiation temperature. Finally, $C$ is given by, 
\begin{equation}
C=\frac{\left[1+K \Lambda_{2 s 1 s} n_H x_{1s} \right]}{\left[1+K \Lambda_{2 s 1 s} 
n_H x_{1s}+K \beta_B n_H  x_{1s} \right]}.
\end{equation}
where, $\Lambda_{2 s 1s} \simeq 8.22 ~{\rm s}^{-1}$ is the decay rate of the 
metastable $2 s$ level to the $1s$ level via two-photon emission, $n_H x_{1s}$ is 
the number density of neutral ground state $H$ atoms, the fraction of hydrogen atoms in 
the $1s$ state is given by $x_{1s}$ and $K=\frac{\lambda_{\alpha}^3}{8 \pi H(z)}$, 
where $H(z)$ is the Hubble expansion rate at redshift $z$ and $\lambda_\alpha$ 
is the wavelength of the Ly-$\alpha$ transition from the $2 p$ level to the $1 s$ 
level. Note that around the recombination epoch, as the fraction of hydrogen 
atoms in the $2s$ or higher states are negligible compared to those in the $1s$ 
state, to a good accuracy, $x_{1s} = 1 - x_e$. As described in 
ref.~\cite{Peebles:1968ja}, this factor $C$ corresponds to the probability of the 
transition from $n=2 $ state to the ground state before getting photoionized. 

Finally, $I_X(z)$, which describes the effect due to annihilating DM particles, is given 
by $I_X(z) = I_{X_i}(z) +  I_{X_{\alpha}}(z)$ \cite{Galli09, Galli:2013dna}. Here $I_{X_i}(z)$ denotes the 
contribution from direct ionization from the ground state of the hydrogen atom, 
thus increasing $x_e$ with time. $I_{X_{\alpha}}$ denotes the contribution to the 
increment of $x_e$ due to ionization from the $n=2$ state. Note that the factor 
$(1-C(z))$ denotes the respective ionization probability. The respective expressions are given by,
\begin{align*}
 I_{X_i}(z) &= - \frac{\chi_i(z)}{(1 + z) H(z) n_H(z) E_i} \left( \frac{dE}{dV dt} \right), \\
 I_{X_{\alpha}}(z) &= - \frac{(1 - C(z)) \chi_\alpha(z)}{(1 + z) H(z) n_H(z) E_\alpha} \left(\frac{dE}{dV dt} \right) .
\end{align*}
In the expressions above, $\left( \frac{dE}{dV dt} \right)$ refers to the energy 
injection from the annihilation of DM, as given by Eq.~[\ref{eq:dmann}], 
$\chi_i \simeq (1- x_e)/3 \simeq \chi_{\alpha}$, which denote the fractions 
of the injected energy contributing towards ionization and excitation, respectively. 
The rate of collisional 
excitation of hydrogen ($1s \to 2s, 2p$) due to DM annihilation is similar to that 
of direct ionization, and can be estimated by replacing $E_i$ by the Lyman-$\alpha$ 
energy $E_\alpha$ and $\chi_i(z)$ by the fraction $\chi_\alpha(z)$ of the injected 
energy going into excitations. As mentioned, once an atom is in the $n = 2$ state, 
there is a probability $(1 - C(z))$ for it to be ionized by CMBR photons. 
 
The enhancement of the free electron fraction around the recombination leads to a broadening of the last scattering surface. Thus, the temperature and polarization power spectrum are modified. While for the TT power spectrum, the effect of annihilation generally suppresses the power at all scales, larger $x_e$ and the increased width of the last scattering surface enhance the polarization power 
spectra at large scales \cite{Padmanabhan:2005es}. Thus, the annihilation of DM during the recombination epoch can have a significant impact  \cite{Kawasaki:2021etm, Slatyer:2015jla, evoli2013cosmic,Padmanabhan:2005es, Chluba:2005uz}. 

The  constraints from Planck collaboration on the parameter $\text{P}_{ann} \leq 1.9 \times 10^{-7} {\rm m}^3\,{\rm sec^{-1}\,Kg^{-1}}$ 
(95\% upper bound)\cite{Planck:2018vyg}, under the assumption that DM only annihilates 
into a pair of electron-positron \cite{Planck:2018vyg}. Including high multipoles ($l$) \texttt{Plik lite} 
likelihood the limit is relaxed to $P_{\rm ann} \leq 2.1 \times 10^{-7} ~{\rm m}^3 \,
\rm sec^{-1}\,Kg^{-1}$ (95\% upper bound). In this work, high-$l$ $l$ TTTEEE lite, 
low-$l$ TT, low-$l$ EE, lensing data set has been used, as it reduces the 
computational time. However, the constraints can be relaxed by up to a factor 
of 2 \cite{Planck:2019nip}. Including Planck lensing and BAO data further 
tightens the constraints. For comparison, it is noteworthy that the correct relic 
abundance of DM via thermal freeze-out requires a thermally averaged 
DM annihilation cross-section in the ballpark of $\langle \sigma |v| \rangle 
\approx 2 \times 10^{-26} ~{\rm cm}^{3}\,\rm{s^{-1}}$ \cite{Planck:2018vyg}.
Apart from affecting the recombination process, the annihilation of DM in the early 
Universe can also affect the growth of perturbations \cite{Bertschinger:2006nq}. We have 
checked the same for some benchmark scenarios, as mentioned in Sec.~\ref{sec:Likelihood analysis and Posteriors}. 
In the present work, we have not included this effect, as it generally leads to
weak constraints on the DM annihilation cross-section. Further, DM annihilation 
into the SM particles at lower redshift affects the reionization process \cite{Mapelli:2006ej}. We 
will not include such effects in the present work. 
\subsection{DM scattering with SM}
\label{sec:DM scattering with SM}
As mentioned in the introduction, the presence of DM-electron interaction 
terms in the lagrangian leads to both DM annihilation into a pair of $\bar{e}e$, 
and DM-$e$ scattering. In this subsection, we review the effect of DM-$e$ scattering 
in the early Universe. The scattering between DM and $e$ in the early Universe,
around and before the recombination epoch, leads to the exchange of energy and 
momentum to the DM from the SM plasma. Thus, DM clumps less at the small 
scales, and generally, the growth of density perturbation on small length scales is suppressed. Consequently, the temperature power spectrum of the CMBR is reduced, 
especially in the small length scales \cite{xu18,Dvorkin13,kim18,Boddy:2018wzy,Sigurdson:2004zp}.  

Following refs. \cite{Dvorkin13, kim18,Boddy:2018wzy,kim21,Buen-Abad:2021mvc,Boddy:2022tyt}, we discuss the 
implications of DM-$e$ drag quantitatively in this context. The presence of 
DM-electron scattering leads to drag between the SM and DM fluids. Around 
the recombination epoch, which is of interest in the present context, both DM 
and baryons are non-relativistic. We denote the (non-relativistic) velocity of DM 
and electron by $\vec{v}_{\chi}$ and $\vec{v}_e$ respectively, and the respective 
momenta scale as $a^{-1}$, where $a$ denotes the scale factor at the same 
epoch. The DM particles, around the epoch of 
recombination ($T_r \lesssim 1$ eV), were not in thermal equilibrium with the 
SM plasma, while the photons and baryons maintain the same temperature thanks 
to the efficient Thomson scattering, as discussed in the previous section. We denote 
the temperature of the baryon fluid by $T_b$, the respective phase space distribution 
function for the electrons are denoted by $f_e(v_e)$. Further, we parametrize the (unperturbed) velocity distribution function of the DM fluid $f_{\chi}(v_{\chi})$ by an effective temperature $T_{\chi}$ \footnote{For a thermal relic DM candidate, $T_{\chi}$ can be simply obtained from the decoupling temperature $T_D$, and is given by $T_{\chi}(z) = \dfrac{T_D a_D^2}{a(z)^2}$, where $a_D$ and $a(z)$ denote the scale factors at the DM decoupling epoch and at the redshift $z$ 
respectively.} 

In the following, we review the DM-electron drag following ref. \cite{kim21}. 
The drag force per unit mass of the DM, $d\vec{v}_\chi / dt$, can be obtained by  
estimating the momentum transfer (per unit time) to the DM particle in a scattering 
with an electron and subsequently taking an average over the respective velocity 
distribution functions. The change in DM momentum $(\Delta \vec{p}_\chi)$ per 
collision is, to leading order in the (non-relativistic) velocities,  given by,
\begin{equation}
\Delta \vec{p}_\chi = \frac{m_\chi m_e}{m_\chi + m_e}|\vec{v}_\chi - \vec{v}_e|(\hat{n} - \frac{\vec{v}_\chi - \vec{v}_e}{|\vec{v}_\chi - \vec{v}_e|})
\end{equation}
where $\hat{n}$ is the direction of the scattered DM particle in the center-of-mass 
frame and $m_e$ and $m_\chi$ denote the mass of the electron and DM 
respectively. The acceleration experienced by the DM is then 
\begin{equation}
\frac{d \vec{v}_\chi}{d t} = \frac{\rho_e}{m_\chi + m_e} \int dv_e v_e^2 f_e(v_e) \int \frac{d\hat{n}_e}{4\pi} \int d\hat{n}\left(\frac{d\sigma(|\vec{v}_{\chi} - \vec{v}_e|)}{d\hat{n}}\right)|\vec{v}_{\chi} - \vec{v}_e|^2(\hat{n} - \frac{\vec{v}_{\chi} - \vec{v}_e}{|\vec{v}_{\chi} - \vec{v}_e|})
\end{equation}
where $\rho_e$ is the energy density of the free electrons. Here $d\sigma(v)/d\hat{n}$ 
is the differential cross-section for electron-DM scattering in the center-of-mass frame, 
which is a function of the respective relative velocity $\vec{v} = \vec{v}_{\chi}- \vec{v}_e$  
between the DM particle and electron and $\hat{n}_e=\frac{\vec{v_e}}{v_e}$. The cross-section is parametrized as a function 
of the relative speed $v$ as follows,  
\begin{equation}
\bar{\sigma}(v)=\sigma_{\rm drag} v_r^n, 
\end{equation}
where $n$ is an integer. In the present study, we will consider velocity-independent 
drag term, i.e. $n=0$, see e.g. \cite{Buen-Abad:2021mvc}. Note that in the literature 
generally, the DM particles are assumed to have a Maxwell-Boltzmann (MB) velocity 
distribution, characterized by the ``temperature'' $T_{\chi}$\footnote{This 
description is accurate if the self-interaction rate of $\chi$ particles is significant 
(i.e., comparable or larger than the Hubble expansion rate). Also, in the case of 
thermal relic $\chi$ particles, one may use effectively $T_{\chi}(z <z_D)= T_{\chi}(z_D) \dfrac{a(z_D)^2}{a(z)^2}$, where $z_D$ denotes the redshift corresponding 
to thermal freeze-out of $\chi$, and $a(z)$ denotes the scale factor at redshift $z$.}.  
Considering both $\chi$ and $e$ follows MB velocity distribution, their relative 
velocity $\vec{v_r} = \vec{v}_{\chi}- \vec{v}_{e}$ also follows the same, with a 
mean relative velocity given by the difference of the respective mean velocities, 
and a variance $\left(\dfrac{T_b}{m_e} + \dfrac{T_{\chi}}{m_{\chi}}\right)$ per direction
\footnote{It has been argued that, for negative $n$, e.g. $n = -2$ and $n= -4$,  
the assumption of MB distribution for DM particles $\chi$ can lead  to a 
discrepancy in the heat exchange rate by a factor of 2-3 compared to a more 
accurate treatment using the Fokker-Planck approximation for the respective 
collision operator  \cite{Ali-Haimoud:2018dvo}. Note that such negative power 
of $v$ can appear if DM candidate possesses electric dipole moment, or for millicharged DM \cite{Dubovsky:2001tr,Dubovsky:2003yn,Sigurdson:2004zp,Dolgov:2013una}. }.
Further, as the present analysis involves very early Universe, the peculiar velocity 
has been assumed to be sufficiently small compared to the respective thermal velocities 
and the dispersion in the respective relative velocities. Thus, non-linear terms involving 
the peculiar velocity have been neglected in estimating the drag force \cite{Dvorkin13}, 
in the case of velocity-independent drag term, i.e., for $\bar{\sigma}(v) = \sigma_{\rm drag}$.  
For an improved approach, see e.g. ref.~\cite{Boddy:2018wzy}. A detailed discussion on 
the validity can be found in ref.~\cite{Ali-Haimoud:2023pbi}.

\subsubsection{Cosmological perturbations with DM - $e$ scattering }
\label{sec:DM-e-scattering}
In this subsection, we discuss the cosmological perturbation  
equations governing the density and velocity perturbations of DM and baryons in 
the presence of DM-$e$ interaction\footnote{Although we only consider DM-$e$ 
scattering, as the $e$, protons (and Helium ions) interact appreciably, the effect of 
such interaction can be treated similarly as that of DM-proton scattering \cite{kim21,Buen-Abad:2021mvc}.}. 
We adopt the synchronous gauge and follow the notations of ref.~\cite{Ma:1995ey}. In the synchronous gauge, when DM interaction is absent, $\theta_\chi$ takes a zero solution and can therefore be ignored. However, this is not possible in the presence of interactions.
The density contrasts $\delta_\chi$ and $\delta_b$ and divergence 
of (bulk) velocity $\theta_\chi$ and $\theta_b$ of DM and baryons in the Fourier space, 
respectively, evolve following the differential equations \cite{Ma:1995ey,Dvorkin13,Xu:2018efh,Gluscevic:2017ywp} :  
\begin{align}
\dot{\delta_\chi} & =-\theta_\chi-\frac{\dot{h}}{2}, \quad \dot{\delta_b}=-\theta_b-\frac{\dot{h}}{2}, \\
\dot{\theta_\chi} & =-\frac{\dot{a}}{a} \theta_\chi+c_\chi^2 k^2 \delta_\chi+R_\chi\left(\theta_b-\theta_\chi\right) \label{eq:deltachi}, \\
\dot{\theta_b} & =-\frac{\dot{a}}{a} \theta_b+c_b^2 k^2 \delta_b+R_\gamma\left(\theta_\gamma-\theta_b\right)+\frac{\rho_\chi}{\rho_b} R_\chi\left(\theta_\chi-\theta_b\right) \label{eq:deltab}.
\end{align}
In the above equations, $\rho_{\chi},~ \rho_{b}$ are the respective energy densities and $c_{\chi},~ c_{b}$ are the speeds of sound in the DM and 
baryon fluids, respectively. These are given by;
\[
c_{\chi}^2 = \frac{\dot{P}_{\chi}}{\dot{\rho}_{\chi}} = \frac{k_B T_{\chi}}{m_\chi} \left(1 - \frac{1}{3} \frac{d \ln T_{\chi}}{d \ln a}\right) ,
\]
\[
c_b^2 = \frac{\dot{P_b}}{\dot{\rho_b}} = \frac{k_B T_b}{\mu} \left( 1 - \frac{1}{3} \frac{d \ln T_b}{d \ln a} \right),
\]
where $\mu$ is the mean molecular weight of the baryonic fluid \cite{Ma:1995ey}. The 
overdot represents a derivative with respect to conformal time, $k$ is the wave 
number of a given Fourier mode, $a$ is the scale factor and $h$ is the trace of the 
spatial part of the metric 
perturbation. The sound speeds $c_{\chi}$, and $c_b$ are generally different, since, 
the interaction of DM with the SM particles, even if non-vanishing, is assumed to be 
rather small. Further, $\delta_{\chi, b}$ and 
$\theta_{\chi, b}$ are the density fluctuations and velocity divergences 
respectively, of the fluids in Fourier space. In addition, the scattering 
between $\chi$ and $e$ leads to heat exchange between DM and baryon 
fluid. This modifies the evolution of the respective temperature $T_b$ and 
$T_{\chi}$ \cite{kim21,Buen-Abad:2021mvc},
\footnote{Similar expressions have been obtained in the context of DM-proton 
scattering in the literature \cite{kim18,Dvorkin13}.}
\begin{align}
\label{eq:dTdzdrag}
-(1 + z) H(z) \frac{dT_b}{dz} + 2 H(z) T_b & =  2 \frac{\mu}{m_e} R_\gamma (T_r - T_b) 
+ 2 \frac{\mu}{m_\chi} R_\chi' (T_\chi - T_b), \\ 
\label{eq:dTdz}
-(1 + z) H(z) \frac{dT_\chi}{dz} + 2 H(z) T_\chi & =  2 R_\chi'(T_b - T_\chi).  
\end{align}
where $T_r$ is the radiation (photon) temperature and 
$\mu$ is the mean molecular weight. The terms proportional to $R_\gamma$ and $R_\chi$ in 
Eqs.~[\ref{eq:deltab}], ~[\ref{eq:deltachi}], ~[\ref{eq:dTdzdrag}] describe the momentum 
transfer between interacting fluids, acting as a drag force between the fluids. The 
momentum-transfer rate coefficient $R_\gamma$ arises from Compton scattering 
between photons and electrons. The rate coefficient for DM-electron scattering is derived in Sec.~\ref{subsubsec:Drag} following the ref. \cite{Boddy:2018wzy} given by, 
\begin{equation}
R_\chi=a \rho_e \frac{\mathcal{N}_n \sigma_{\rm drag}}{m_\chi+m_e}\left(\frac{T_\chi}{m_\chi}+\frac{T_b}{m_e} + \frac{V^2_{\text{rms}}}{3}\right)^{(n+1) / 2},
\label{eq:rchi}
\end{equation} 
also, 
\begin{equation}
R_\gamma=a \frac{4\rho_\gamma}{3\rho_b} n_e \sigma_T, 
\label{eq:rchi}
\end{equation} 
where $\sigma_T$ is the Thomson scattering cross-section, $V^2_{\rm rms} = \langle \vec{V}^2_{\chi} \rangle_{\xi} =
\int \frac{dk}{k} \, \Delta_{\xi} \left( \frac{\theta_b - \theta_c}{k} \right)^2$ and $\langle \dots \rangle_{\xi}$ denotes an average with respect to the primordial curvature perturbation, and  
$\Delta_{\xi} \simeq 2.4 \times 10^{-9}$ is the primordial curvature variance per $\log k$~\cite{PhysRevD.82.083520},
$\sigma_{\rm drag}$ denotes the DM-electron scattering cross-section, $\mathcal{N}_n \equiv 2^{(5+n) / 2} \Gamma(3+n / 2) /(3 \sqrt{\pi})$, and $\rho_e=(1-Y_{\mathrm{He}}) \rho_b x_e m_e / m_p$ is the electron density, 
$Y_{\mathrm{He}}$ is the helium mass fraction, $m_p$ is the proton mass and $x_e$ is the ionization fraction. The heat-transfer rate coefficient is $R_\chi^{\prime}=R_\chi m_\chi /\left(m_\chi+m_e\right)$. 

\newpage
In the present context, we focus on the case where $n = 0$. In this scenario, the relative bulk velocity between the DM and baryon fluids is small compared to the thermal relative velocity between the particles, allowing us to neglect the $V_{\text{rms}}$ term.
However, for the cases where $n = -2$ and $n = -4$, the DM scattering rate is weak in the early Universe, allowing the relative bulk velocity to exceed the thermal velocity. In such cases, the $V_{\text{rms}}$ term cannot be neglected relative to the thermal velocity, and the full expression must be used.

\section{DM-electron Interaction: An Effective Operator Approach}\label{sec:pert2}
In this section, we discuss relevant interactions of DM with $e$ considered in 
the present work. In the following discussion, we assume that such interactions are 
mediated by very heavy mediators with the mediator mass much larger than twice 
the mass of DM (i.e. 2$m_{\chi}$). Therefore, we describe the relevant interaction 
terms using effective operators with the following form,
\begin{equation}
\mathcal{L}_{\text{eff}} = g_\text{{eff}} \, \mathcal{O}_{\chi} \, \mathcal{O}_e.
\label{eq:Leff}
\end{equation}
In the above equation, $O_{\text{$\chi$}}$ and $O_e$ are operators which are bilinear 
in the DM and electron fields, respectively, and $g_{\rm eff}$ denotes the respective 
effective coupling. It is convenient to denote the product of these two operators as 
$\mathcal{O}^{e}$,  where the superscript $e$ indicates that the operator describes DM-electron interactions; $\mathcal{O}^{e}= \mathcal{O}_{\chi} \, \mathcal{O}_e$. More generally, 
several such effective operators may be present, and in such scenarios, the effective 
lagrangian is given by $\mathcal{L}_\text{eff} = \sum_{i} {g_{\rm eff,i}} \mathcal{O}^{e}_{\rm i}$ 
where $\mathcal{O}^{e}_{i}$ denotes the $i^{th}$ operator and $g_\text{eff,i}$
denotes the corresponding interaction strength, for details see. \cite{Beltran:2008xg,Kumar:2013iva,Brod:2017bsw,Liang:2024ecw}. So far we have used $\chi$ to denote the DM particle in general. In the following, we use $\psi$ and $\phi$ to denote Dirac fermionic DM and real scalar DM, respectively. The symbol $\chi$ will be used in contexts where the discussion applies to both types of DM, as before. In Table~[\ref{tab:effective-operators}], we describe the lowest dimensional effective operators depicting DM-$e$ interactions for (Dirac) fermionic
DM ($\psi$), 
and for real scalar DM ($\phi$). The respective mass dimensions of these operators are 
six and five for fermionic DM and real scalar DM particles, respectively. Considering one 
type of operator at a time, we perform an expansion in the DM-electron relative speed $v_r$, 
and only keep the leading order term for each of these operators. The 
DM-$e$ scattering cross-section for each of these operators, up to the leading order 
in the relative speed $v_r$ between the DM particle and $e$ or the corresponding 
momentum transfer $\boldsymbol{q} $, thus obtained, are mentioned in the third 
column. Finally, in the fifth column, the nature of the (thermally) averaged DM annihilation 
cross-section is mentioned. Note that in this context, $v_{\chi, rel}$ denotes the relative 
speed between the annihilating DM particles. As discussed before, the same effective 
operator can lead to both DM annihilation into a pair of electron and positron, and 
DM-$e$ scattering. However, as the relative velocity $v_{\chi, rel}$ is non-relativistic, 
among the operators considered, only the ones contributing to the $s$-wave annihilation 
can be effective at late times.

\begin{table}[ht]
\centering
\setlength{\tabcolsep}{14pt} 
\resizebox{1.1\textwidth}{!}{
{\Large 
\begin{tabular}{|c|c|c|c|c|}
\hline
\textbf{S.No} & \textbf{Operator ($\mathcal{O}_i^e$)} & \begin{tabular}[c]{@{}c@{}}\textbf{Non-relativistic Reduction} \\ \textbf{(for drag)}\end{tabular} & $\boldsymbol{\sigma_{\rm drag}}$ & $\boldsymbol{\langle \sigma v_{\chi,\text{rel}} \rangle} (s-wave)\approx
$ \\
\hline
$F_1$ & $\bar{\psi} \psi \bar{e} e$ & $4 m_\psi m_e \mathcal{O}_1^e$ & $g_\text{eff}^2 \dfrac{m_\psi^2 m_e^2}{\pi (m_\psi + m_e)^2}$ & No s-wave\\
\hline
$F_2$ & $\bar{\psi} \gamma^5 \psi \bar{e} e$ & $4 m_e \mathcal{O}_2^e$ & $\sim v_r^2$ suppressed &  $\dfrac{g_\text{eff}^2}{2\pi} \sqrt{1 - \dfrac{m_e^2}{m_\psi^2}}(m_\psi^2)$  \\
\hline
$F_3$ & $\bar{\psi} \psi \bar{e} \gamma^5 e$ & $-4 m_\psi \mathcal{O}_3^e$ & $\sim v_r^2$ suppressed & No s-wave \\
\hline
$F_4$ & $\bar{\psi} \gamma^5 \psi \bar{e} \gamma^5 e$ & $-4 \mathcal{O}_4^e$ & $\sim v_r^4$ suppressed &  $\dfrac{g_\text{eff}^2}{2\pi} \sqrt{1 - \dfrac{m_e^2}{m_\psi^2}}(m_\psi^2)$\\ 
\hline
$F_5$ & $\bar{\psi} \gamma^\mu \psi \bar{e} \gamma_\mu e$ & $4 m_\psi m_e \mathcal{O}_1^e$ & $g_\text{eff}^2 \dfrac{m_\psi^2 m_e^2}{\pi (m_\psi + m_e)^2}$ & $\dfrac{g_\text{eff}^2}{2\pi}m_\psi^2 \sqrt{1 - \dfrac{m_e^2}{m_\psi^2}} \left( 2 + \dfrac{m_e^2}{m_\psi^2} \right)$ \\
\hline
$F_6$ & $\bar{\psi} \gamma^\mu \gamma^5 \psi \bar{e} \gamma_\mu e$ & $8 m_\psi m_e (\mathcal{O}_5^e - m_e \mathcal{O}_6^e)$ & $\sim v_r^2$ suppressed & No s-wave \\
\hline
$F_7$ & $\bar{\psi} \gamma^\mu \psi \bar{e} \gamma_\mu \gamma^5 e$ & $-8 m_e (m_\psi \mathcal{O}_7^e + \mathcal{O}_6^e)$ & $\sim v_r^2$ suppressed & $\dfrac{g_\text{eff}^2}{\pi} \sqrt{1 - \dfrac{m_e^2}{m_\psi^2}}m_e^2$ \\
\hline
$F_8$ & $\bar{\psi} \gamma^\mu \gamma^5 \psi \bar{e} \gamma_\mu \gamma^5 e$ & $-16 m_\psi m_e \mathcal{O}_8^e$ & $\dfrac{g_\text{eff}^2}{4} \cdot \dfrac{48 m_e^2}{\pi (1 + m_e/m_\psi)^2}$ & $\dfrac{g_\text{eff}^2 m_\psi^2}{2\pi} \sqrt{1 - \dfrac{m_e^2}{m_\psi^2}} \left( \dfrac{m_e^2}{m_\psi^2} \right)$ \\
\hline
$F_9$ & $\bar{\psi} \sigma^{\mu \nu} \psi \bar{e} \sigma_{\mu \nu} e$ & $32 m_e m_\psi \mathcal{O}_8^e$ & $g_\text{eff}^2 \dfrac{48 m_\psi^2 m_e^2}{\pi (m_\psi + m_e)^2}$ & $\dfrac{g_\text{eff}^2 }{2\pi} m_\psi^2\sqrt{1 - \dfrac{m_e^2}{m_\psi^2}} \left[ 16\left( 1 + \dfrac{m_e^2}{m_\psi^2} \right) \right]$ \\
\hline
$S_1$ & $(\phi^\dagger \phi)(\bar{e} e)$ & $2 m_e \mathrm{I}_e$ & $\dfrac{g_\text{eff}^2}{16 \pi (1 + m_\phi/m_e)^2}$ & $\dfrac{g_\text{eff}^2}{8\pi} \left( 1 - \dfrac{m_e^2}{m_\phi^2} \right)^{3/2}$ \\
\hline
$S_2$ & $(\phi^\dagger \phi)(\bar{e} \gamma^5 e)$ & $2 m_e \mathcal{O}_9^e$ & $\sim v_r^2$ suppressed & $\dfrac{g_\text{eff}^2}{8 \pi} \sqrt{1 - \dfrac{m_e^2}{m_\phi^2}}$ \\
\hline
\end{tabular}
}
}
\caption{ Summary of the effective operators,  matrix elements, and cross-sections for DM-electron scattering, and thermally averaged DM annihilation cross-sections \cite{Beltran:2008xg,Buckley:2011kk,Kumar:2013iva,Brod:2017bsw,Liang:2024ecw,Fitzpatrick:2010em}.}
\label{tab:effective-operators}
\end{table}

In Table~[\ref{tab:effective-operators}] the operators 
$\mathcal{O}^{e}_i$ are, thus, given by :  
\begin{align*}
\mathcal{O}_1^e & \equiv  \xi^\dagger_s \xi_s \xi^\dagger_r \xi_r, \\
\mathcal{O}_2^e & \equiv i \boldsymbol{s}_\psi \cdot \boldsymbol{q}, \\
\mathcal{O}_3^e & \equiv i \boldsymbol{s}_e \cdot \boldsymbol{q}, \\
\mathcal{O}_4^e & \equiv (\boldsymbol{s}_\psi \cdot \boldsymbol{q})(\boldsymbol{s}_e \cdot \boldsymbol{q}), \\
\mathcal{O}_5^e & \equiv \boldsymbol{s}_\psi \cdot \boldsymbol{v}_e^{\perp}, \\
\mathcal{O}_6^e & \equiv i \boldsymbol{s}_\psi \cdot (\boldsymbol{s}_e \times \boldsymbol{q}), \\
\mathcal{O}_7^e & \equiv \boldsymbol{s}_e \cdot \boldsymbol{v}_e^{\perp}, \\
\mathcal{O}_8^e & \equiv \boldsymbol{s}_\psi \cdot \boldsymbol{s}_e.\\
\mathcal{O}_{9}^e & \equiv i \boldsymbol{s}_e \cdot (\boldsymbol{q} \times \boldsymbol{v}_e^{\perp}), 
\end{align*}

In the above expressions, $\boldsymbol{s}_\psi =\frac{1}{2} \xi^\dagger_s \vec{\sigma} \xi_s$,  $\boldsymbol{s}_e = \frac{1}{2} \xi^\dagger_r \vec{\sigma} \xi_r$ and $\boldsymbol{q} $ denotes the momentum transfer  defined as $\boldsymbol{q=p_1 - p_2}$ , $\boldsymbol{v}^\perp_e = \frac{\boldsymbol{p}_1 + 
\boldsymbol{p}_2}{2m_\psi} - \frac{\boldsymbol{k}_1 + \boldsymbol{k}_2}{2m_e}$, where $\boldsymbol{p_1}$ and $\boldsymbol{p_2}$ are the momenta of the initial-state DM and $e$ in the scattering, and $\boldsymbol{k_1}$ and $\boldsymbol{k_2}$ are the momenta of the final-state DM and $e$.
In the subsequent sections, we describe the implications of DM annihilation and 
scattering with electrons in the light of CMBR data. 

\section{Results}
\label{sec:result}
The effect of DM annihilation and scattering with electrons in the very early 
Universe, around the epoch of recombination, can be captured by two 
parameters, $P_{\rm ann}$ and $\sigma_{\mathrm{\rm drag}}$ respectively. As the presence of DM-electron interaction terms in the lagrangian leads to both DM annihilation into a pair of electron and positron, and scattering between DM and electron, we extend the six parameters of the base \(\Lambda\)CDM by these two additional parameters \(\sigma_{\rm drag}\) and \(P_{\text{ann}}\). The base parameters are: \(\omega_b\) (baryon density), \(\omega_c\) (dark matter density), \(\theta_s\) (sound horizon angle), \(A_s\) (scalar perturbation amplitude), \(n_s\) (spectral index), and \(\tau_{\text{reio}}\) (reionization optical depth).

We have used a modified version \cite{kim18} of \texttt{CLASS} 
\cite{lesgourgues2011cosmic,Blas_2011} to solve the relevant equations
Eq.~[\ref{eq:deltachi}] and Eq.~[\ref{eq:deltab}] to estimate the TT, EE power 
spectrum. The likelihood for the model parameters is estimated using 
Bayesian statistics with suitable priors. Using Markov Chain Monte 
Carlo (MCMC) method, with the help of \texttt{MontePython} 
\cite{Brinckmann:2018cvx}, then, the maximum likelihood and the 
confidence intervals are determined. In particular, we obtain the 95\% upper 
limit on the parameters $P_{\rm ann}$ and $\sigma_{\rm drag}$. These 
upper limits are then used to obtain constraints on the interaction strengths 
of the effective operators, as mentioned in Table~[\ref{tab:constraints}].  
\footnote{The obtained chains are tested using the Gelman-Rubin criterion 
(\texttt{R-1} convergence criterion). The commonly used threshold is 
\texttt{R-1} < 0.02. However, for the \texttt{lite} dataset in the case of 
annihilation, we observed a change in the 95\% C.L. value even below 
this threshold. Therefore, we used a stringent criterion to ensure convergence, 
\texttt{R-1} < 0.002.} 

Note that, in the present study, the terms involving DM annihilation in the 
relevant perturbation equations have not been considered. We have 
checked that it does not affect the constraints appreciably. Further, we have 
not included the contribution of the drag term in the temperature evolution, 
as described in Eq.~[\ref{eq:dTdzdrag}] for the same reason. 

\subsection{Likelihood analysis and Constraints of Cosmological Parameters}
\label{sec:Likelihood analysis and Posteriors}
In this section, we describe the constraints on the parameters ($P_{\rm ann}$ 
and $\sigma_{\rm drag}$) relevant for describing the DM interactions in the 
light of CMBR data. For the analysis, we have used the Planck 2018  baseline 
data with high-$\ell$ TTTEEE lite, low-$\ell$ TT, low-$\ell$ EE, and lensing dataset.

As mentioned above, we have used a modified version of the publicly available 
code \cite{kim18} \texttt{CLASS} \cite{lesgourgues2011cosmic,Blas_2011}  
together with the MCMC simulator \texttt{MontePython} \cite{Brinckmann:2018cvx} 
to obtain the posterior distributions of the parameters considered in our study. 
We have varied the mass of DM as follows : $m_{\chi}\in \{ {\rm 1 \,MeV,  ~10 \,MeV, ~100 \,MeV, 
~1 \,GeV, ~10 \,GeV}\}$. Further, assuming flat priors, the following upper 
limits on the flat priors for the relevant cosmological parameters, respectively, for 
the DM masses we considered: 
\begin{eqnarray}
    \sigma_{\text{\rm drag}} &=& [1 \times 10^{-24}, 2 \times 10^{-24}, 3 \times 10^{-22},3 \times 10^{-21}, 3 \times 10^{-20}], \\
    \text{$P_{\mathrm{ann}}$} &=& [3 \times 10^{-6}, 2 \times 10^{-6}, 4 \times 10^{-6}, 3 \times 10^{-6}, 3 \times 10^{-6}].
\end{eqnarray}
The respective lower limits are set to $0$.

\begin{table}[!h] 
    \centering
    \begin{tabular}{|c|c|c|c|}
        \hline
        DM mass ($m_\chi$) & $\sigma_{\mathrm{\rm drag}}$ (95\% C.L) & $P_{\mathrm{ann}}$ 
         (95\% C.L) & $\sigma_{\mathrm{\rm drag}}$ ($P_{\mathrm{ann}}$ = 0)(95\% C.L)\\
        \hline
        1 MeV & $1.38 \times 10^{-26}$ & $2.31 \times 10^{-7}$ & $1.22 \times 10^{-26}$\\
        \hline
        10 MeV & $9.75 \times 10^{-26}$ & $2.35 \times 10^{-7}$ & $8.70 \times 10^{-26}$\\
        \hline
        100 MeV & $9.43 \times 10^{-25}$ & $2.38\times 10^{-7}$ & $8.3 \times 10^{-25}$\\
        \hline
        1 GeV & $9.44 \times 10^{-24}$ & $2.406 \times 10^{-7}$ & $8.5 \times 10^{-24}$\\
        \hline
        10 GeV & $10.3 \times 10^{-23}$ & $2.41 \times 10^{-7}$ & $9.44 \times 10^{-23}$\\
        \hline
    \end{tabular}
    \caption{Constraints on $\sigma_{\text{\rm drag}}$ (in cm\textsuperscript{2}) and $P_{\mathrm{ann}}$ (in m \textsuperscript{3}/sec/kg) for different $m_{\chi}$.}\label{tab:constraints}
    \centering
\end{table}

\begin{figure}
    \centering
        \includegraphics[width=.49 \textwidth]{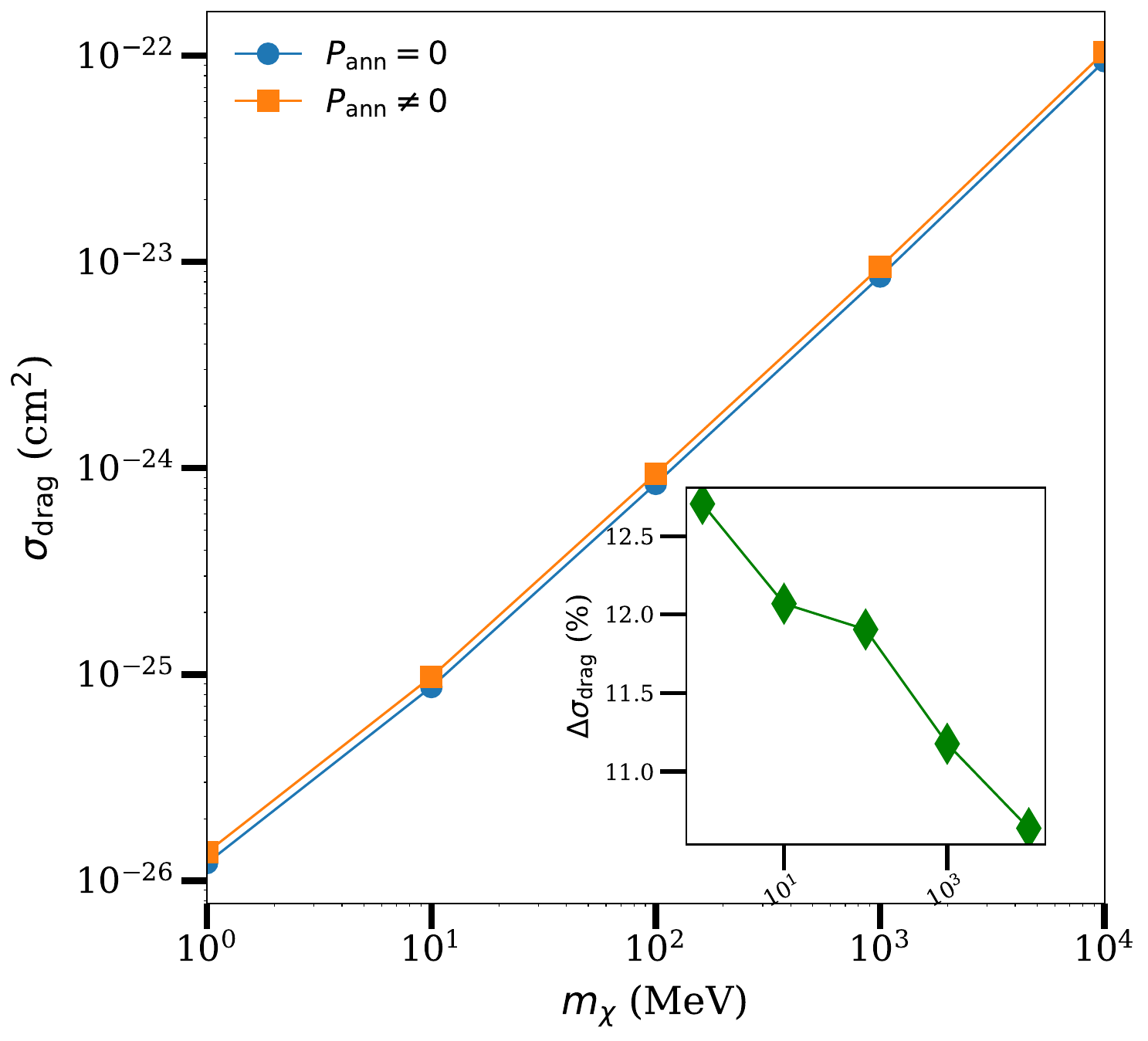}
        \includegraphics[width=.48\textwidth]{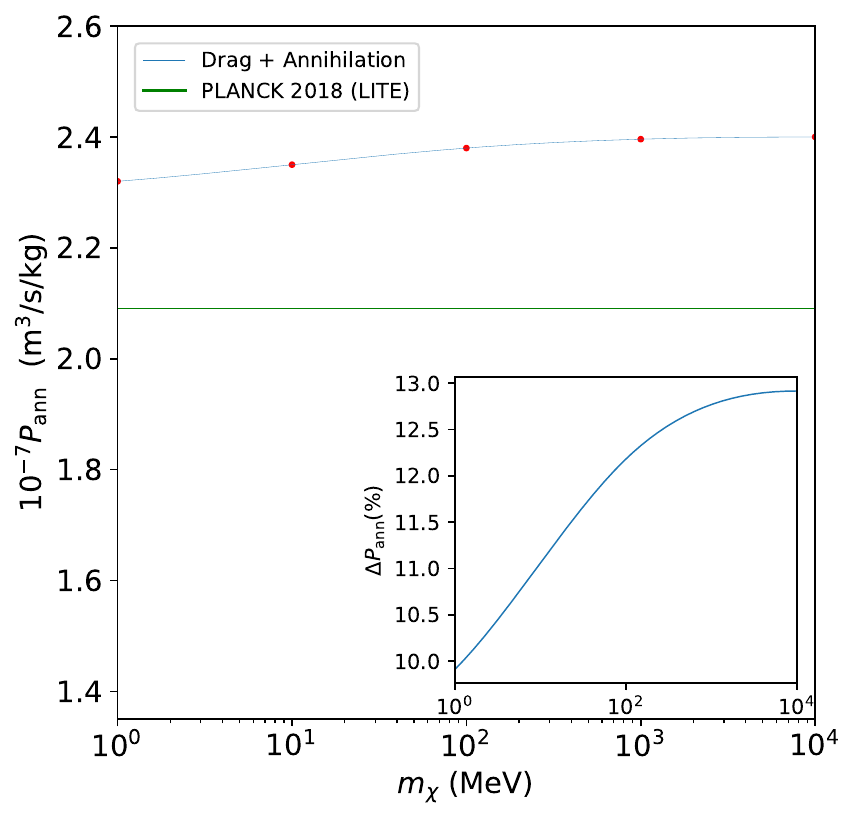}
        \caption{In the left panel shows the 95\% upper bound on $\sigma_{\rm drag}$ for various dark matter masses, assuming $P_{\rm ann} = 0$ (in blue), using the high-$\ell$ TTTEEE lite, low-$\ell$ TT, low-$\ell$ EE, and lensing dataset from \rm{PLANCK 2018}, along with the corresponding upper bound on $\sigma_{\rm drag}$ when both annihilation and drag effects are considered (in green). The bottom right panel illustrates the percentage difference between these two values.
        The right panel presents the 95\% upper bound on $P_{\rm ann}$ as a function of $m_\chi$ using the same data set as mentioned above. The solid green line indicates the constraint on $P_{\rm ann}$ ($\leq 2.1 \times 10^{-7} ~{\rm m}^3 \, \rm sec^{-1}\,Kg^{-1}$) obtained from the same dataset in the absence of DM-electron scattering.}
    \label{fig:combined}
\end{figure}

In Table~[\ref{tab:results1mev}], the results of the statistical analysis have been 
mentioned for $m_\chi$ = 100 MeV. In particular, the best-fit values and the limits on 
the relevant parameters ($\Omega_b,~ \Omega_{\rm drag},~ \theta_s$, $A_s,~ n_s$ 
and $\tau_{\rm reio}$) (at 1$\sigma$ and at 2-$\sigma$) have been shown. For 
different $m_{\chi}$, as described above, the results of the statistical analysis are 
presented in Table~[\ref{tab:constraints}]. In particular, in this table, the limits 
(at 95\% C.L.) on $\sigma_{\rm drag} {\rm (cm^2)}$ and $P_{\rm ann} {\rm (cm^{3}/sec/kg)}$ are mentioned for the different choices of $m_{\chi}$. In the right panel of 
Fig.~[\ref{fig:combined}], the 95\% upper limits on the parameter $P_{\rm ann}$ for 
different masses of DM ($m_{\chi}$) have been shown. The solid line depicts the 
annihilation parameter $P_{\rm ann}$ in the absence of the drag term obtained 
using the same dataset, as mentioned above. Note that, in the absence of the drag 
term, the upper limit on $P_{\rm ann}$ is independent of the mass of DM $m_{\chi}$. 
However, as shown in the right panel of Fig.~[\ref{fig:combined}], when DM annihilation 
into $\bar{e}e$, as well as DM-electron scattering effects are considered, the constraints 
on the parameter $P_{\rm ann}$ shows moderate dependence of $m_{\chi}$. The 
variation is less than $\mathcal{O}(10)\%$, considering $m_{\chi}$ 
within the range 1 MeV to 10 GeV. This can be inferred from Eq.~[\ref{eq:deltachi}] 
and Eq.~[\ref{eq:rchi}], as for a fixed $P_{\rm ann}$ and $\sigma_{\rm drag}$, 
the parameter $R_{\chi}$ varies with $m_{\chi}$. Consequently, the best fit and 
the 95\% upper limits on $P_{\rm ann}$ (and $\sigma_{\rm drag}$) are dependent 
on $m_{\chi}$. Further, this figure shows that the constraints on $P_{\rm ann}$ 
are relaxed by approximately 10-15\% in the presence of $\sigma_{\rm drag}$ 
as compared to the case when only $P_{\rm ann}$ is considered, i.e. 
$\sigma_{\rm drag}$ is set to 0. 

In the left panel of Fig. [\ref{fig:combined}], 95\% C.L. upper limits on 
$\sigma_{\rm drag}$ have been shown for various $m_{\chi}$. 
For smaller $m_{\chi}$, the number density 
of the DM is large, and the probability of the scattering events are higher. The 
effect of DM-electron scattering, for a given cross-section $\sigma_{\rm drag}$, 
would be more prominent for lighter DM. This can lead to an increase in the DM 
flux incident on an electron and, thus, enhance the number of DM-electron scattering 
events. \footnote{ In addition, the relative (thermal) speed of light DM particles, for 
a fixed $T_{\chi}$, would be higher. Although around the epoch of recombination, 
$T_b$ is generally significantly larger than $T_{\chi}$, and thus, the thermal velocity 
of the electrons are dominant.} Consequently, as shown in the left panel of 
Fig. [\ref{fig:combined}], the upper limits on $\sigma_{\rm drag}$ is relaxed  
as the mass of DM is increased. Also, the same figure shows that the bound 
on the scattering cross-section ($\sigma_{\rm drag}$) is relaxed for the entire 
range of $m_{\chi}$, in the presence of DM annihilation. The percentage change, 
given by  $\Delta \sigma_{\rm drag} \% = \frac{\sigma_{\rm drag} (P_{\rm ann} \neq 0) - 
\sigma_{\rm drag} (P_{\rm ann} = 0)}{\sigma_{\rm drag} (P_{\rm ann} = 0)} 
\times 100$, is about $10\%$.

\begin{figure}
    \centering 
        \includegraphics[width=.49\textwidth]{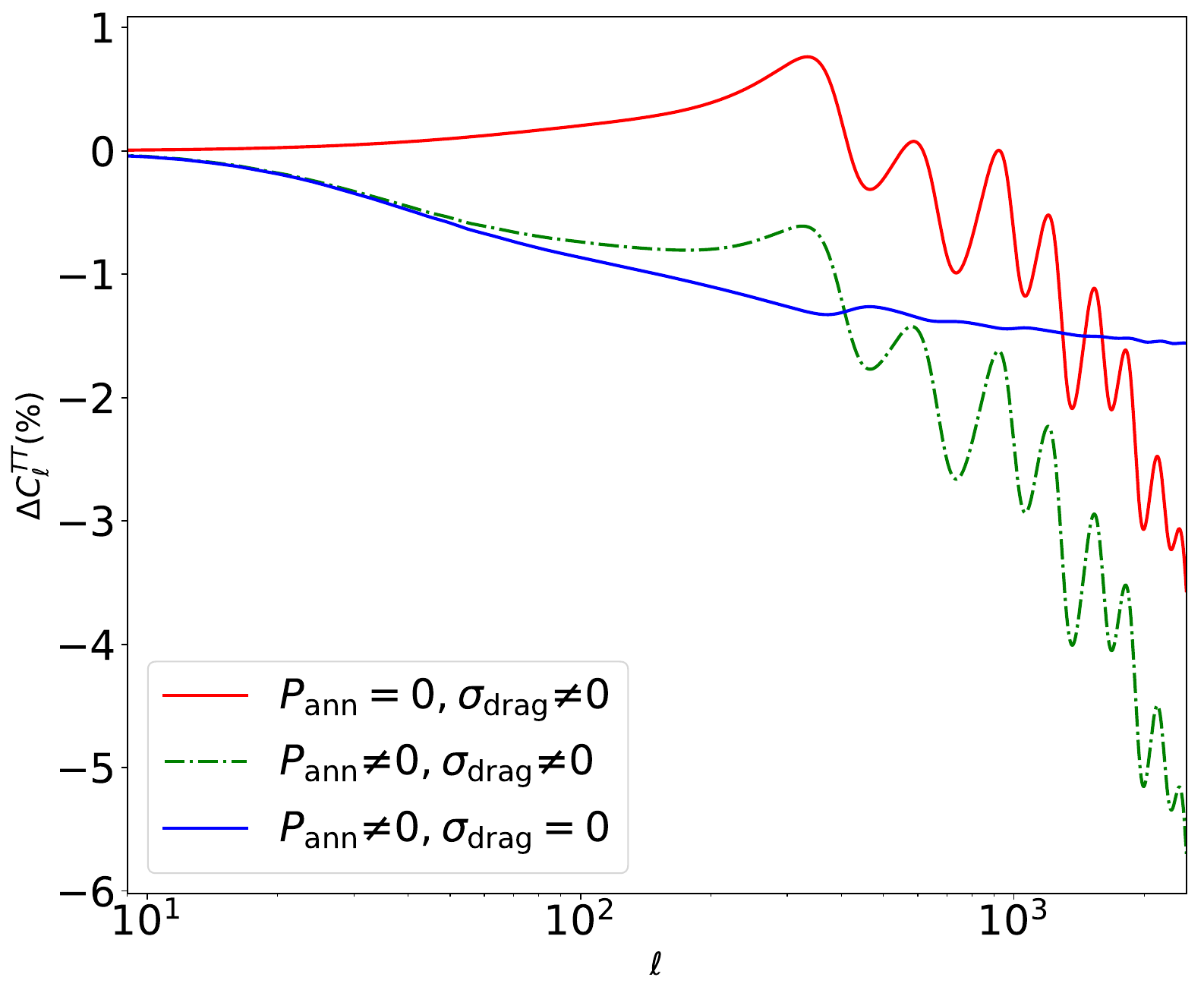}
        \includegraphics[width=.50\textwidth]{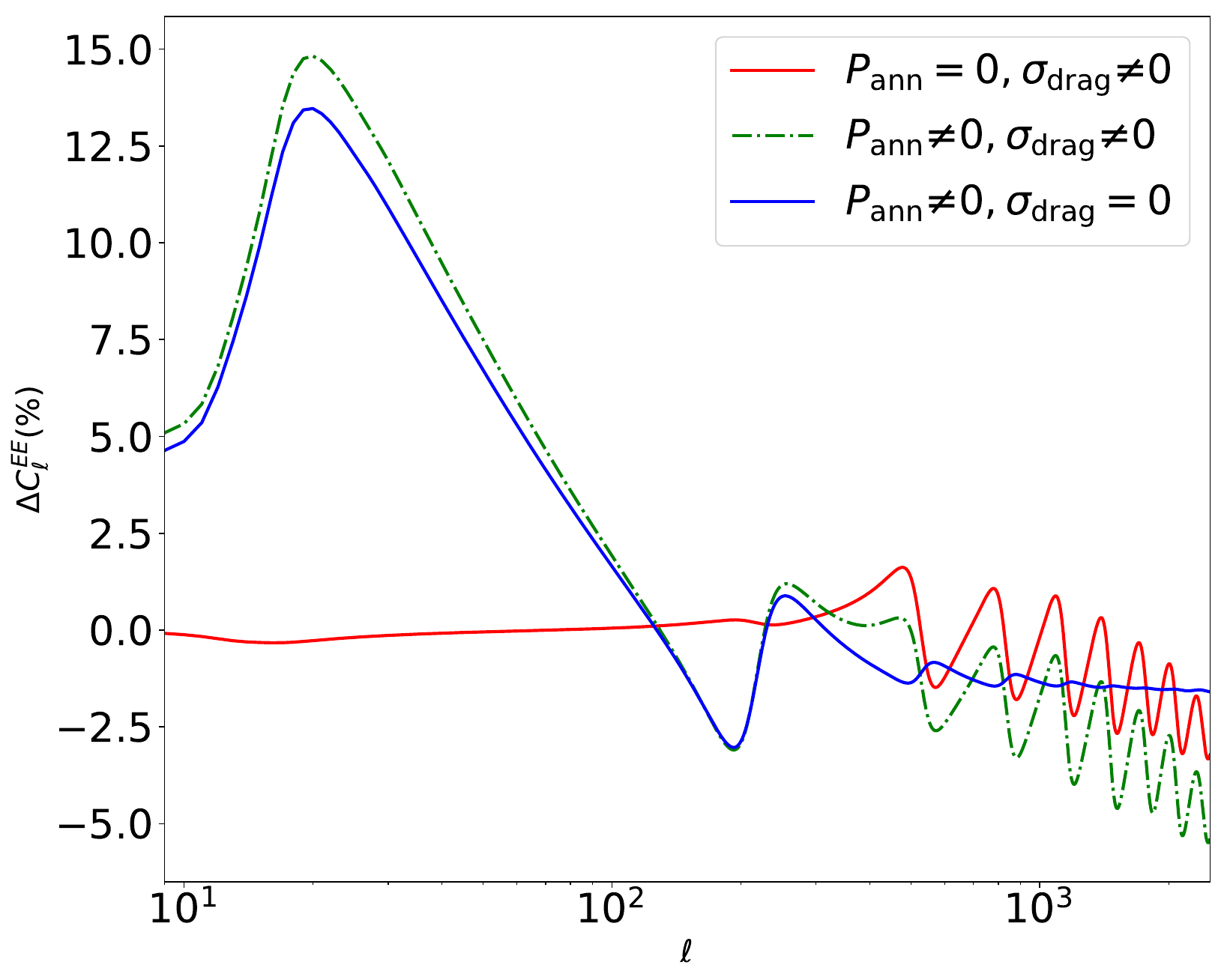}
        \caption{Comparison of the effect of DM-Electron scattering with annihilation on the TT power 
        spectrum for DM mass $m_\chi = 100$ MeV. For each curve, we fixed $\sigma_{\text{\rm drag}} 
        = 9.44 \times 10^{-25} \, \text{cm}^2$ and $P_{\text{ann}} = 2.38 \, \text{m}^3/\text{s}/\text{kg}$ 
        to its corresponding 95\% C.L. value obtained with CMBR data.}
    \label{fig:ClTTEE}
\end{figure}

In order to further understand the effect of the DM interactions on the 
CMBR anisotropy, in Fig.~[\ref{fig:ClTTEE}] we present the percentage 
deviation in the power spectrum of the temperature anisotropy ($C_l^{TT}$) 
and E-mode polarization anistropy ($C_l^{EE}$) as a function of the 
multipole moment $l$ due to DM interactions for $m_{\chi} = 100$ MeV. 
In the left panel, $\Delta C_l^{TT} = \dfrac{C_{l \rm c}^{TT}- C_{l}^{TT}}{C_{l}^{TT}} 
\times 100$ has been shown, where $C_{l \rm c}^{TT}$ and $C_l^{TT}$ denote the 
respective power of the two-point correlations of temperature anisotropy at multipole 
$l$ with and without DM interactions, respectively. Similarly, the left panel shows 
$\Delta C_l^{EE} =  \dfrac{C_{l\rm c}^{EE}- C_{l}^{EE}}{C_{l}^{EE}} \times 100$, 
where $C_{l \rm c}^{EE}$ and $C_l^{EE}$ denote the respective power of the 
two-point correlations of E-mode polarization anisotropy at multipole $l$ with 
and without DM interactions, respectively. In this figure, the best-fit values for
the cosmological parameters in the $\Lambda$CDM have been used, and for 
the additional parameters $\sigma_{\rm drag}$ and $P_{\rm ann}$, the 95\% 
upper limit, as described in Table~[\ref{tab:constraints}] have been used. 

Note that the presence of the drag term with electrons contributes to the increase 
of the rate of change of $\theta_{\chi}$ and thus, adversely affects the growth of 
density contrast of DM, $\delta_{\chi}$. This effect is particularly prominent in the 
smaller length scales, (thanks to a sizable $c_{\chi}$) i.e. for larger multipole moments 
$l$. This leads to a depletion in the temperature power spectrum for higher multipole 
moments $l$, as shown in the left panel of Fig.~[\ref{fig:ClTTEE}]. A similar trend 
is observed for the polarization power spectrum for larger multipole moments, as 
demonstrated in the right panel of the same figure. 
Further, in the presence of the DM-electron scattering term, DM is dragged by baryons, 
as baryons undergo oscillations around the recombination epoch. This affects the 
baryon loading, enhancing the power spectrum in the large length scales, (i.e. $k$ is 
sufficiently small) \cite{Chen:2002yh,Hu:2008hd,kim18}, which leads to the rise in the 
temperature correlation $\Delta C_l^{TT}$.  As demonstrated in the left panel of 
Fig.~[\ref{fig:ClTTEE}], the temperature power spectrum is enhanced for smaller 
multipoles ($l \lesssim 320$). 

As discussed previously,  the effect of DM annihilation into $\bar{e}e$ 
raises the baryon temperature $T_b$ (as compared to the scenario, where such 
annihilation is absent) in addition to increasing the free electron fraction $x_e$ around 
the last scattering surface. The consequent increase in the optical depth, in turn, 
depletes the temperature power spectrum at all length scales.   Thus, both drag and 
annihilation show depletion of the CMBR temperature power spectrum in a similar way for large multipole 
moments $l$. In the left panel and in the right panel of Fig. [\ref{fig:ClTTEE}], these 
effects are demonstrated. However, in presence of DM annihilation into $\bar{e}e$, 
as the free electron fraction $x_e$ is enhanced, the polarization (E mode) 
power spectrum shows a substantial increase for large length scales 
($2\lesssim l\lesssim 200$), as demonstrated in Fig. [\ref{fig:ClTTEE}]. 

\begin{figure}
    \centering
    \includegraphics[width=0.9\textwidth]{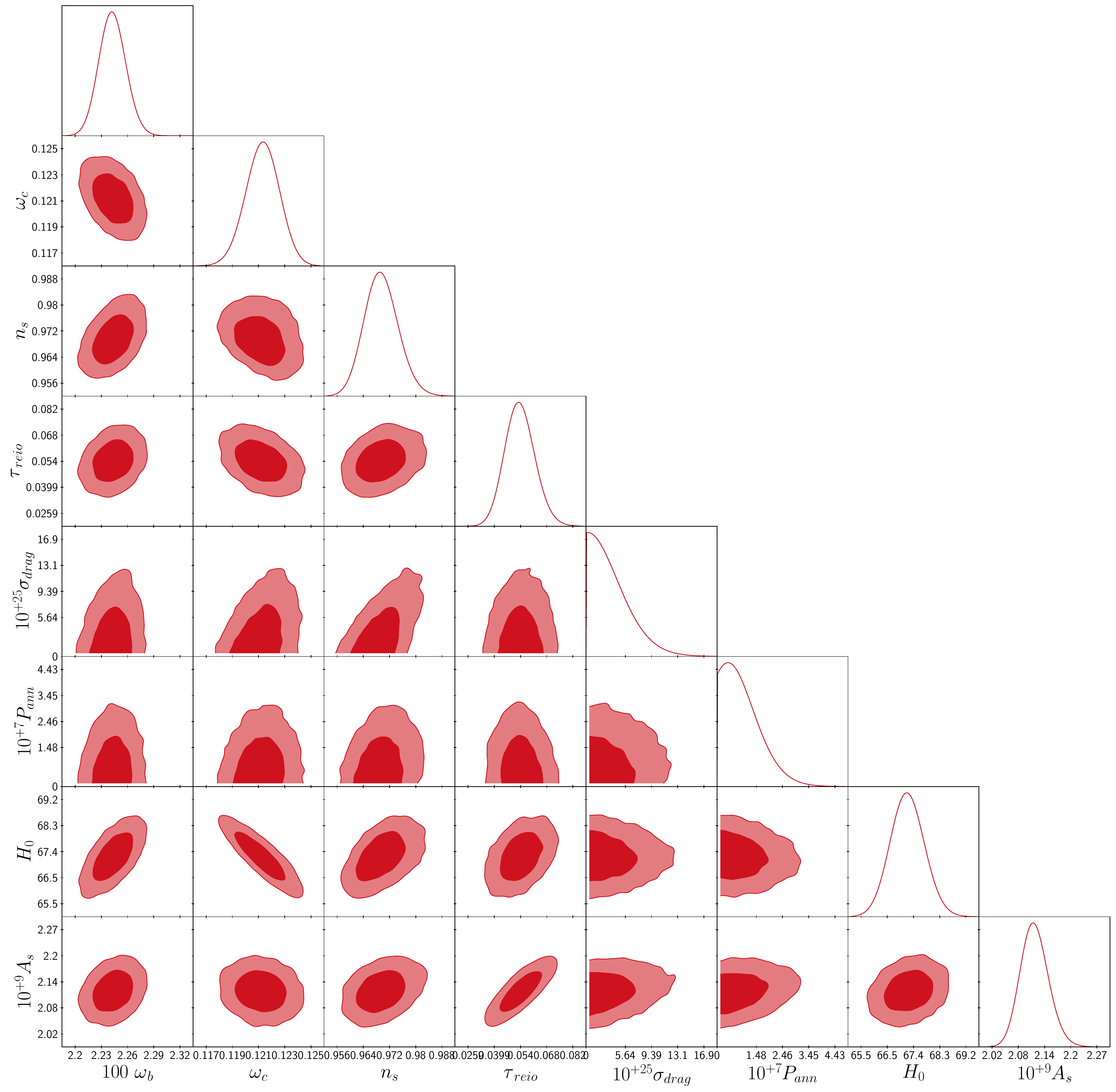}
    \caption{1-d posterior distribution  of 6+2 parameter model with parameters 
    $\omega_b$, $\omega_{\rm drag}$, $\theta_s$, $A_s$, $n_s$, $\tau_{reio}$, 
    $\sigma_{\rm drag}$, $P_{\rm ann}$ for $m_\chi$ = 100 MeV. }
    \label{fig:tr1}
\end{figure}

From the above discussion, it is evident that, although in the small length scales 
the effects of drag and annihilation can lead to depletion of power, their effects 
can be very different in the larger length scales. Consequently, negligible 
correlation is observed in the respective posterior distributions, which is 
shown in Fig. [\ref{fig:tr1}] for $m_{\chi} = 100$ MeV.

\begin{figure}
    \centering
    \includegraphics[width=0.70\textwidth]{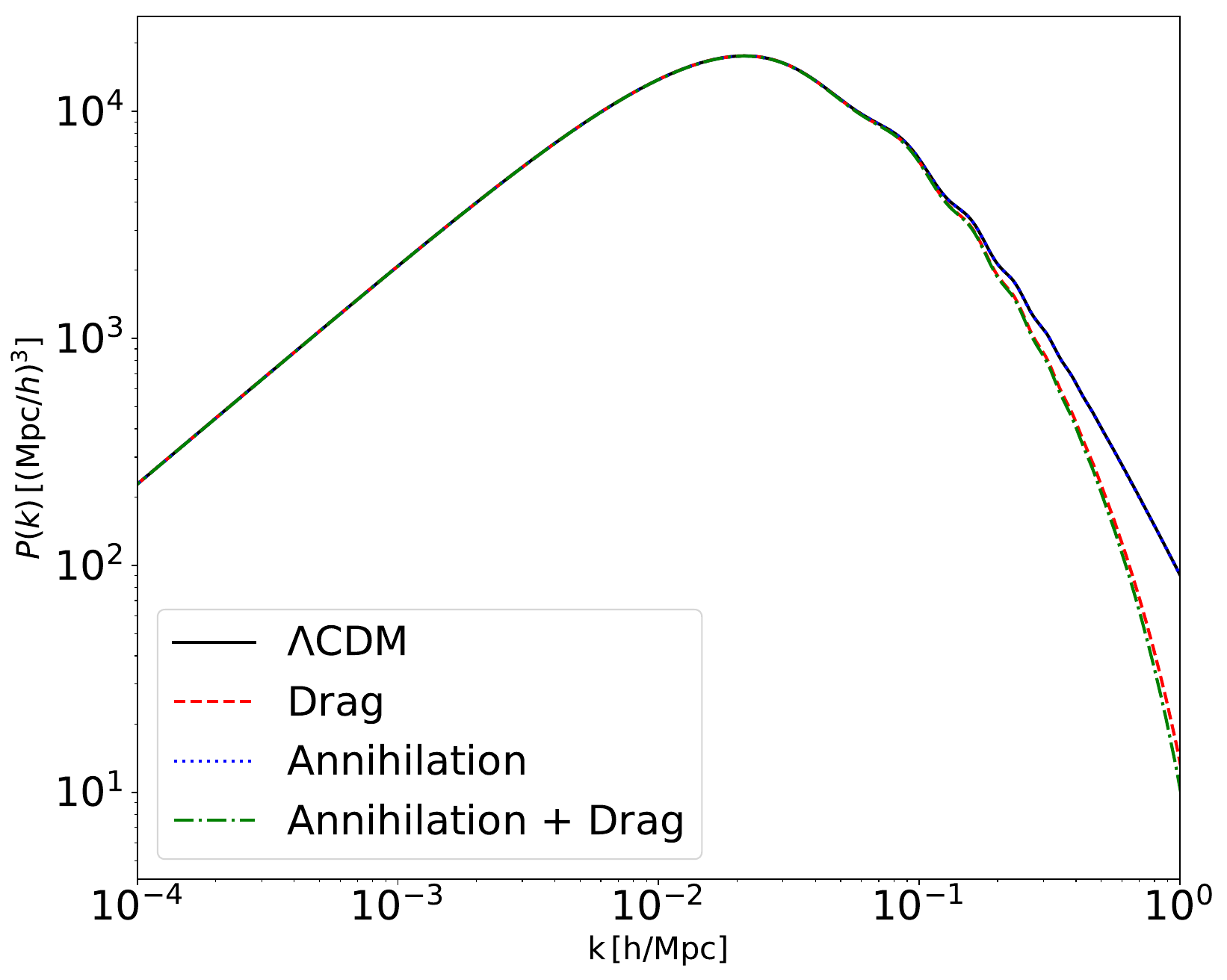}
    \caption{The effect of DM-electron scattering with annihilation on the matter power spectrum for DM mass  \( m_\chi = 100 \, \text{MeV} \). For each curve, we fixed \( \sigma_{\rm {drag}} = 9.44 \times 10^{-25} \, \text{cm}^2 \) and \( P_{\rm ann} = 2.38 \, \text{m}^3/\text{s}/\text{kg} \) to its corresponding 95\% C.L. value obtained with CMBR data.. The lines for Annihilation and $\Lambda$CDM overlap at high $k$ (small length scales), with differences appearing only at very low $k$ (large length scales), where there is a small bulge around $10^{-5}$.} 
    \label{fig:TTpk}
\end{figure}

As discussed, for heavier $m_{\chi}$, the number of scattering 
events decreases for the same cross-section $\sigma_{\rm drag}$. Consequently, 
the upper limit on the scattering cross-section $\sigma_{\rm drag}$ is enhanced to 
achieve a similar fit to the CMBR anisotropy data, which is demonstrated in the left 
panel of Fig.~[\ref{fig:combined}]. The depletion of the temperature power spectrum 
at large $k$ is generally slightly reduced (for the 95\% C.L. upper limit on 
$\sigma_{\rm drag}$ for larger $m_{\chi}$.  In the small $k$ region, similar 
to the case of a lighter DM, a small rise in the temperature power is observed. 
The 95\% upper limit on $P_{\rm ann}$  is somewhat relaxed. We find that 
the upper limit on $P_{\rm ann}$ is in agreement with the limit obtained without 
considering the drag term with electrons to about $\lesssim 10\%$, as shown in 
the same figure. A depletion of power in the small scales can also be seen from 
the matter power spectrum, as shown in Fig.~[\ref{fig:TTpk}], when the DM-electron 
drag term is considered. The corresponding large $k$ modes enter the horizon 
earlier in the radiation-dominated Universe.

Note that using Planck data (Planck TT, TE, EE+lowE+lensing), 
the constraints on the DM annihilation parameter $P_{\rm ann}$, 
for s-wave annihilation, is given by $2.33  \times 10^{-28} \, \text{cm}^3 \, \text{s}^{-1} \, \text{GeV}^{-1}$ or equivalently, 
$1.9 \times 10^{-7} \, \text{m}^3 \, \text{s}^{-1} \, \text{kg}^{-1}$ at 95\% confidence level
\cite{Planck:2018vyg}. \footnote{As mentioned, in this work, we have used the 
\texttt{Plik lite} data set in our analysis. We have checked that  the upper limit 
on $P_{\rm ann}$, as obtained with $\sigma_{\rm drag} =0$, differs if the full 
Planck data set is used instead.} The upper limit on this parameter is independent 
of the mass of annihilating DM $m_{\chi}$, and is also independent of specific 
annihilation channels for annihilation into the standard model particles 
(except neutrinos). On the averaged annihilation cross-section $\langle \sigma v \rangle$, for a fixed $m_{\chi}$, the strongest constraints are obtained for DM annihilation 
into $\bar{e}e$.  This is because, in this scenario, the injected energy thermalizes 
with the SM plasma almost instantly, and consequently, the fraction of the energy 
absorbed remains large and is independent of the redshift around the epoch of 
recombination (i.e., $f_{\rm eff}(z) \simeq f_{\text{eff}}$)
\cite{Slatyer:2015jla,Planck:2018vyg,Lopez-Honorez:2013cua}. We obtain the value of $f_{\rm eff}$ for different DM masses from refs.~\cite{Lopez-Honorez:2013cua,Slatyer:2015jla,Planck:2018vyg}. The constraint on $P_{\rm ann}$ has profound implications for the thermal production of DM particles in the early Universe, particularly via s-wave annihilation into SM particles. The corresponding lower bounds on $m_{\chi}$ range from $m_{\chi} \geq 9 \, \text{GeV}$ for annihilation into $\tau^+ \tau^-$ pairs, up to $m_{\chi} \geq 30 \, \text{GeV}$ for annihilation into $\bar{e}e$ pairs. Assuming thermal DM production, the 95\% confidence level (CL) lower bound on the DM mass is $m_{\chi} \geq 40$ GeV\footnote{For the full Planck dataset, which attempted to constrain $P_{\rm ann}$ without invoking the drag term, the s-wave annihilation cross-section into the $\bar{e}e$ channel, required to satisfy the correct thermal relic abundance, is disfavored for $m_{\chi} \leq 30$ GeV~\cite{Planck:2018vyg}.} for annihilation into $\bar{e}e$ pairs using the \texttt{lite} dataset. When both scattering and annihilation processes are considered together, the constraint on the annihilation cross-section is relaxed, lowering the bound on the DM mass for thermally produced DM in the early Universe to $m_{\chi} \geq 32$ GeV. This trend is expected to hold for the full Planck dataset, though the numerical values could change by a few percent.  
 
 %
 
The correlations of various cosmological parameters, in Fig.~[\ref{fig:tr1}], 
the 1-$\sigma$ and 2-$\sigma$ posterior contours have been shown for 
$m_{\chi} = 100$ MeV. For different values of $m_{\chi}$ we considered, 
the nature of these contours and correlations among different parameters demonstrates a similar trend. We particularly focus on the correlations with various cosmological parameters with the parameters in the DM sector, i.e. 
$\omega_c (\Omega_c = \omega_c \text{h}^2)$, $P_{\rm ann}$ and $\sigma_{\rm drag}$. 

Fig.~[\ref{fig:tr1}] demonstrates that $\sigma_{\rm drag}$ and $P_{\rm ann}$ 
are constrained by the CMBR data. The respective best-fit values are smaller than 
the upper limit by an order of magnitude. Further, the contours terminate at the 
lower limit on the prior (i.e. $0$). Thus, setting  $\sigma_{\rm drag}$ and 
$P_{\rm{ann}}$ to $0$ (i.e. the base $\Lambda$CDM) is consistent with 
the data. Note that no correlation between these two parameters is observed. 
However, for $\sigma_{\rm {drag}}$, we observe noticeable positive correlation 
with $n_s$. This may be understood from the fact that a larger 
$\sigma_{\rm drag}$ prohibits the growth of the DM density contrast, especially at small length scales. In the left panel of Fig.~[\ref{fig:ClTTEE}],  the suppression in the TT spectrum 
at small length scales (corresponding to the higher multipole moments), in the 
presence of the drag term, has been shown. This may be partly compensated 
by raising the scalar spectral index 
$n_s$, which enhances the primordial scalar power at these scales. There is a 
small positive correlation observed between $\omega_{c}$ and $\sigma_{\rm drag}$. 
This may be understood as follows: increasing $\sigma_{\rm drag}$ leads to depletion in the 
growth of the DM density contrast. This may be partly compensated by also increasing 
the DM abundance, thus enhancing $\omega_c$. Further, similar to the $\Lambda$CDM 
model (with only gravitationally interacting  DM) sizable negative correlation 
between the Hubble parameter at the present epoch $H_0$ and $\omega_c$  is also 
observed. This is understood as increasing $\omega_c$ enhances the distance to
the first peak. Consequently, as the angle $\theta_s$ is tightly constrained by the observation, $H_0$ is decreased. There is no noticeable (negative) correlation between $n_s$ and $P_{\rm ann}$. Note that the polarization data can remove any degeneracy in this case \cite{Padmanabhan:2005es}. Finally, a strong negative correlation is observed between the parameters $\sigma_8$ (which depicts the matter power at 8 Mpc), and $\sigma_{\rm drag}$. In the following subsection, in the context of $\sigma_8$ tension, we discuss this in some detail.  

\begin{table}
\setlength{\tabcolsep}{12pt} 
\renewcommand{\arraystretch}{1.5}
\centering
\begin{tabular}{|l|c|c|c|c|} 
 \hline 
Param & best-fit & mean$\pm\sigma$ & 95\% lower & 95\% upper \\ \hline 
$100~\omega{}_{b }$ &$2.238$ & $2.241_{-0.015}^{+0.016}$ & $2.211$ & $2.272$ \\ 
$\omega_c$ &$0.1209$ & $0.121_{-0.0014}^{+0.0013}$ & $0.1184$ & $0.1237$ \\ 
$n_{s }$ &$0.9656$ & $0.9695_{-0.0054}^{+0.0046}$ & $0.9595$ & $0.9798$ \\ 
$\tau{}_{reio }$ &$0.05328$ & $0.05411_{-0.0077}^{+0.0074}$ & $0.0387$ & $0.06949$ \\ 
$10^{+25}\sigma_{\rm drag}$ &$-$ & $3.667_{-3.7}^{+0.91}$ & $-$ & $9.449$ \\ 
$10^{+7}P_{{\rm ann} }$ &$-$ & $0.9729_{-0.97}^{+0.25}$ & $-$ & $2.38$ \\ 
$H_{0 }$ &$67.08$ & $67.19_{-0.56}^{+0.56}$ & $66.07$ & $68.29$ \\ 
$10^{+9}A_{s }$ &$2.106$ & $2.121_{-0.034}^{+0.032}$ & $2.054$ & $2.185$ \\ 
\hline 
\multicolumn{5}{|c|}{$-\ln{\cal L}_\mathrm{min} =506.548$,  $\chi_{\rm min}^2=1013.0960$} \\ 
\hline
\end{tabular} 
\caption{Statistical result of 6+2 parameter model with parameters $\omega_b$, 
$\omega_{cdm}$,$H_0$, $A_s$, $n_s$, $\tau_{reio}$, $\sigma_{\rm drag}$,  
$P_{\rm ann}$ for $m_\chi$ = 100 MeV. } \label{tab:results1mev}
\centering
\end{table}
In general, it is possible that multiple operators contribute to DM annihilation and 
drag interactions, and different operators may dominate each of these processes. For example, the 
operator $F_1$ can contribute significantly to the drag effect, while $F_2$, as given in 
Table~[\ref{tab:effective-operators}], can contribute dominantly to the annihilation process. 
\footnote{Further, it may be possible that DM annihilates into $e\bar{e}$, while the 
drag with the SM plasma is dominated by DM-proton interactions, especially for 
$m_{\chi} \lesssim m_p$, where $m_p$ denotes the mass of the proton.}
In the present context, we take an agnostic approach and treat these processes independently. 

As mentioned, our primary focus is on a velocity-independent drag cross-section, corresponding 
to $n = 0$ in Eq.~[\ref{eq:rchi}]. The effect of DM-electron (and also DM-proton) scattering
with velocity-dependent scattering cross-sections (i.e., $\bar{\sigma}(v_r) = \sigma_{\rm drag} 
v_r^n$ for $(n = -2, n = -4)$, in the absence of DM annihilation (i.e. with $P_{\rm ann}=0$), 
have been previously studied in refs.~\cite{Dubovsky:2001tr,Dubovsky:2003yn,McDermott:2010pa,Dolgov:2013una,Boddy:2018wzy,essig21}
In the following, we briefly discuss such a possibility in the presence of $s$-wave annihilation 
of DM into $e\bar{e}$. For instance, in the case of $n = -4$, the upper limit on $P_{\rm ann}$ at the 95\% C.L. is $2.3 \times 10^{-7} \, \text{m}^3 \, \text{s}^{-1}$ for a DM mass of 1 MeV and 
$2.2 \times 10^{-7} \, \text{m}^3 \, \text{s}^{-1}$ for 1 GeV. Thus, we find that the constraints 
on $P_{\rm ann}$ vary with $m_{\chi}$ in this scenario. Further, the upper limits on DM-electron 
drag parameter $(\sigma_{\rm drag})$, in the presence of DM annihilation, are generally relaxed 
by $\mathcal{O}(10)\%$, depending on the choice of $m_{\chi}$. For $n = -2$ and $n = -4$, the DM 
an improved treatment of the peculiar velocity has been followed, as it may not generally be small 
compared to the thermal velocity dispersion during the relevant epochs ~\cite{Dvorkin13, Xu:2018efh,Ali-Haimoud:2023pbi}.

\subsubsection{Comment on $\text{S}_8$ or $\sigma_{8}$ Tension}
\begin{figure}
    \centering
    \includegraphics[width=0.5\textwidth]{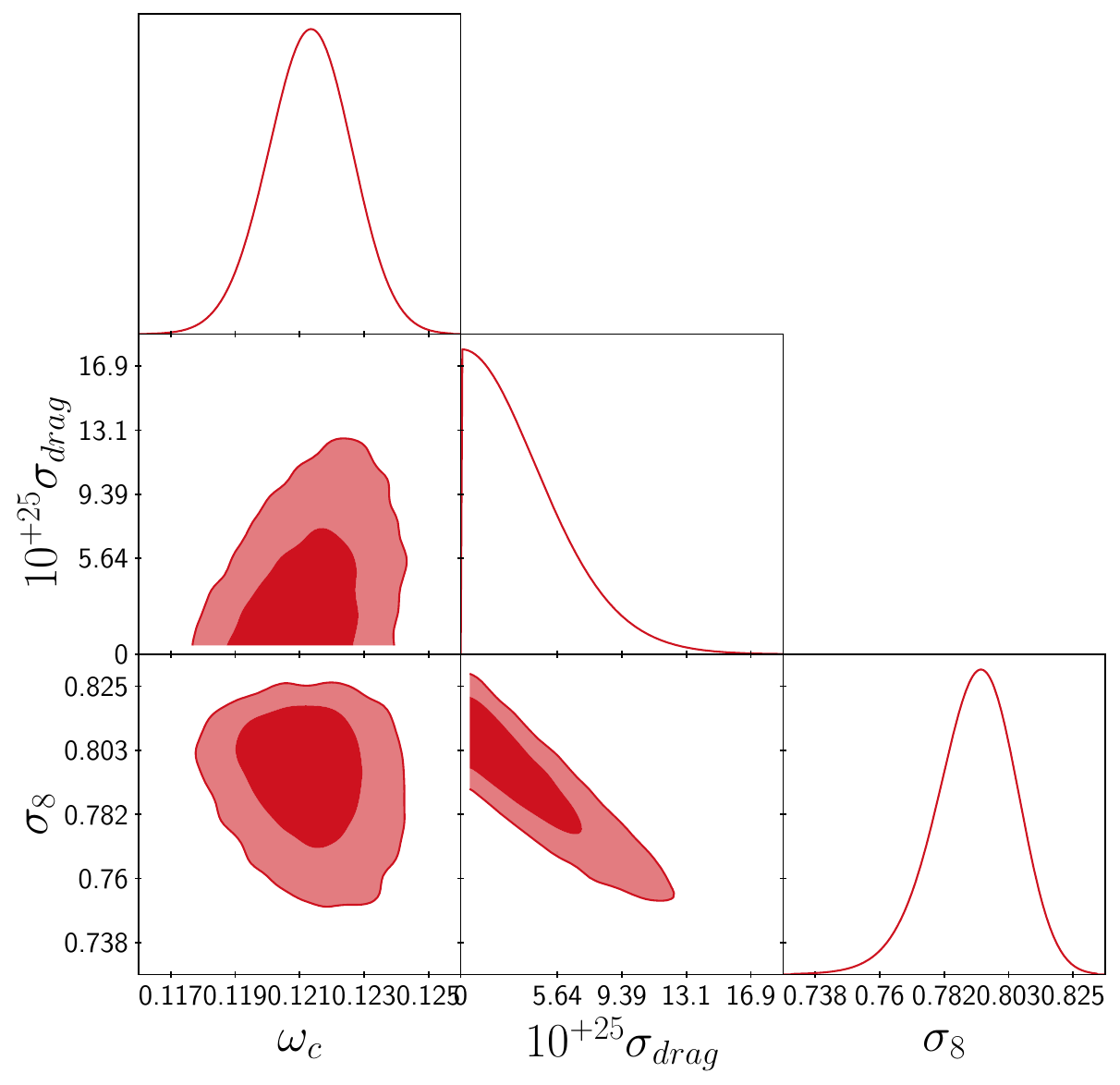}
    \caption{Posterior distribution among $\omega_c$, $\sigma_{\rm drag}$ and $\sigma_8$ for $m_\chi$ = 100 MeV.} 
    \label{fig:sigma8}
\end{figure}
The $S_8$ parameter is defined as
\begin{equation}
    S_8 \equiv \sigma_8 \left(\frac{\Omega_m^0}{0.3}\right)^{0.5},
\end{equation}
where $\Omega_m^0 \equiv \rho^0_m / \rho^0_{cr}$ is the density of matter today as a fraction of the critical density $\rho^0_{cr}$, and $\sigma_8$ measures the rms amplitude of linear matter density fluctuations over a sphere of radius $R = 8 \, \text{Mpc}/h$ at $z = 0$:

\begin{equation}
    \sigma_8^2 = \frac{1}{2\pi^2} \int \frac{dk}{k} W^2(kR) k^3 P(k),
\end{equation}
where $P(k)$ is the linear matter power spectrum today and $W(kR)$ is a spherical top-hat filter of 
radius $R = 8 \, \text{Mpc}/h$.
\begin{table}[H]
\setlength{\tabcolsep}{1pt} 
\renewcommand{\arraystretch}{0.5}
\centering
\begin{tabular}{|l|c|c|c|c|} 
 \hline 
Param & best-fit & mean$\pm\sigma$ & 95\% lower & 95\% upper \\ \hline 
$\sigma_8$ &$0.8102$ & $0.7942_{-0.011}^{+0.02}$ & $0.7611$ & $0.8209$ \\  
\hline 
 \end{tabular} \\  
\caption{Results for $\sigma_8$ with $\sigma_{\rm drag}$ for DM of mass 100 MeV.} \label{tab:sigma8}
\centering
\end{table}

There is a moderate tension emerging between $S_8$ (or $\sigma_8$) as measured from late-Universe 
datasets \cite{KiDS:2020suj,DES:2021bvc} and as indirectly inferred by the CMBR 
\cite{Planck:2018vyg,Planck:2015fie}, i.e., by constraining the $\Lambda$CDM parameters from the CMBR 
and calculating the resulting $S_8$ (or $\sigma_8$). In particular, weak lensing surveys, such as 
Kilo Degree Survey (KiDS) measures $S_8 = 0.759 \pm 0.024$ 
\cite{KiDS:2020suj,Hildebrandt:2020rno,Busch:2022pcx}, and clustering surveys like Baryon 
Oscillation Spectroscopic Survey (BOSS) also consistently finds low $S_8$ (or $\sigma_8$) 
\cite{Philcox:2021kcw,Chen:2022jzq,Zhai:2022yyk,Zhang:2021yna}. Dark Energy Survey (DES) measures 
$S_8 = 0.776 \pm 0.017$ ($\sigma_8 = 0.733 \pm 0.0060$) \cite{DES:2021wwk,DES:2021bvc} (a combined 
analysis of the clustering of foreground galaxies and lensing of background galaxies ). These numbers 
should be compared to the indirect CMBR constraint of $S_8 = 0.834 \pm 0.016$ ($\sigma_8 = 0.811 \pm 
0.0060$) from Planck and ground-based experiments eg, Atacama Cosmology Telescope (ACT), and  South 
Pole Telescope (SPT)\cite{Planck,ACT:2020gnv,Wu:2019hek}.

In the presence of DM-e interaction, we observe a negative correlation between 
the parameter $\sigma_8$ and the scattering parameter $\sigma_{\rm drag}$. This result is 
consistent with ref.~\cite{He:2023dbn} where DM-proton drag was considered. A non-vanishing $\sigma_{\rm drag}$ 
leads to a lower value of $\sigma_8$ compared to the standard $\Lambda$CDM model, 
potentially alleviating the $\sigma_8$ tension. This suggests that interactions in the 
dark sector can play a role in the resolution of the tension. However, a more detailed 
investigation using late-time datasets (such as \rm{KiDS}), is necessary for a robust 
conclusion. Note that the numbers presented in Table [\ref{tab:sigma8}] 
are only representative, and is obtained for $m_{\chi} = 100$ MeV. In this case, for 
$\sigma_{\rm drag}$, we obtain a best fit of $6.452 \times 10^{-26} {\rm cm}^2$, 
which is an order of magnitude below the 95\% upper bound on the same parameter. 
Also, there is no significant improvement in the goodness of fit. As we will discuss 
Subsequently, in this mass range, constraints from the direct detection experiments 
are stringent and generally rule out such large scattering cross-sections. However, 
we checked that, for lighter $m_{\chi} \lesssim 0.5$ MeV, where the direct detection 
constraints are very weak or even absent, the presence of $\sigma_{\rm drag}$ leads 
to a lower $\sigma_8$.

\subsection{Constraints on lagrangian Parameters} 
\label{subsec:constraintsL}

\begin{figure}\label{fig:Leff}
    \centering
    \includegraphics[width=1\textwidth]{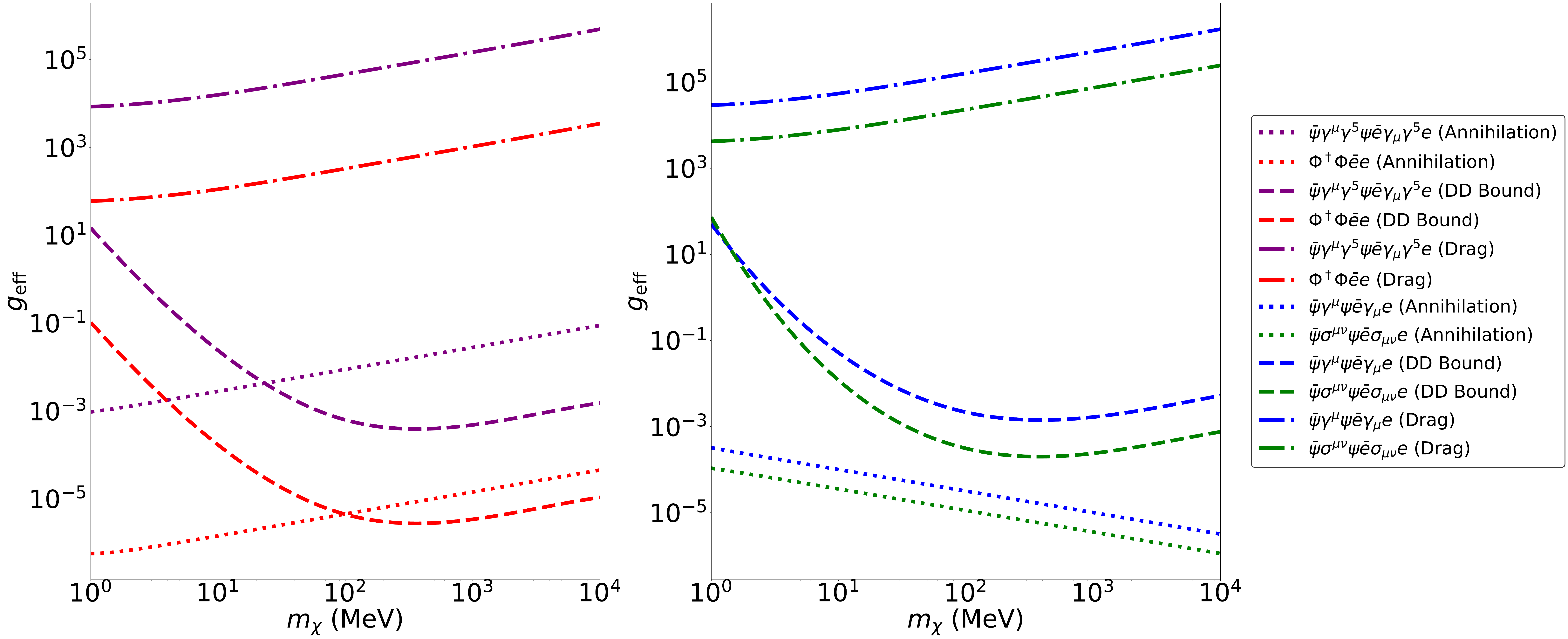}
    \caption{Constraints on the effective coupling $g_{\text{eff}}$ as a function of the mass of DM $m_{\chi}$ has been shown, where the coupling $g_{\text{eff}}$ is expressed in units of $(\mathrm{GeV})^{-2}$ for fermionic DM and $(\mathrm{GeV})^{-1}$ for scalar DM. The dotted (dash-dotted) lines represent constraints from DM annihilation (drag), while the dashed lines indicate limits from direct detection  \cite{SENSEI:2019ibb,XENON:2021qze,SENSEI:2023zdf,DAMIC-M:2023gxo,DarkSide:2022knj,PandaX-II:2021nsg}. Different colors represent vector ($F_5$), pseudo-vector ($F_8$), tensor ($F_9$), and scalar ($S_1$) operators, as defined in Table~\ref{tab:effective-operators}.}\label{fig:g_eff}
    \end{figure}
In this section, we describe the implications of the constraints on $\sigma_{\rm drag}$ and $P_{\rm ann}$ for the lagrangian parameters describing the effective DM-electron interactions. 
Note that when DM annihilation into $e\bar{e}$ is kinematically viable, the upper limits on 
both $\sigma_{\rm drag}$ and $P_{\rm ann}$ can lead to constraints on $g_{\rm eff}$. However, 
generally, more than one effective operator can be present, and different operators may 
contribute to the (velocity-independent) annihilation and drag. Further, the annihilation 
and drag with two different species may be present. We find that the constraints 
on an effective interaction strength $g_{\rm eff}$, as derived from the upper limit on 
$P_{\rm ann}$ is generally stronger as compared to the constraint on the same parameter 
derived from $\sigma_{\rm drag}$. Consequently, the upper limits on $P_{\rm ann}$ for 
velocity-independent annihilation processes, as obtained in the presence of a drag term 
(i.e. keeping both $P_{\rm ann}$ and $\sigma_{\rm ann}$ as independent parameters) have 
been used to constrain the respective lagrangian parameters describing DM-$e$ interactions.

\begin{table}[ht]
\centering
\begin{tabular}{c c c c c}
\toprule
{$m_{\chi}$ (\si{\mega\electronvolt})} & 
{Axial-Vector ($F_8$) ($g_{\mathrm{eff}}$)} & 
{Vector ($F_5$) ($g_{\mathrm{eff}}$)} & 
{Tensor ($F_9$) ($g_{\mathrm{eff}}$)} & 
{Scalar ($S_1$) ($g_{\mathrm{eff}}$)} \\
\midrule
1       & \makecell{$9.51\times10^{-4}$\\($8.47\times10^{3}$)} & \makecell{$3.23\times10^{-4}$\\($2.93\times10^{4}$)} & \makecell{$1.08\times10^{-4}$\\($4.23\times10^{3}$)} & \makecell{$5.66\times10^{-7}$\\($6.00\times10^{1}$)} \\
10      & \makecell{$2.79\times10^{-3}$\\($1.57\times10^{4}$)} & \makecell{$1.01\times10^{-4}$\\($5.45\times10^{4}$)} & \makecell{$3.56\times10^{-5}$\\($7.87\times10^{3}$)} & \makecell{$1.43\times10^{-6}$\\($1.11\times10^{2}$)} \\
$10^2$  & \makecell{$8.83\times10^{-3}$\\($4.65\times10^{4}$)} & \makecell{$3.19\times10^{-5}$\\($1.61\times10^{5}$)} & \makecell{$1.13\times10^{-5}$\\($2.32\times10^{4}$)} & \makecell{$4.50\times10^{-6}$\\($3.29\times10^{2}$)} \\
$10^3$  & \makecell{$2.79\times10^{-2}$\\($1.47\times10^{5}$)} & \makecell{$1.01\times10^{-5}$\\($5.13\times10^{5}$)} & \makecell{$3.56\times10^{-6}$\\($7.40\times10^{4}$)} & \makecell{$1.42\times10^{-5}$\\($1.05\times10^{3}$)} \\
$10^4$  & \makecell{$8.83\times10^{-2}$\\($4.93\times10^{5}$)} & \makecell{$3.19\times10^{-6}$\\($1.71\times10^{6}$)} & \makecell{$1.13\times10^{-6}$\\($2.46\times10^{5}$)} & \makecell{$4.50\times10^{-5}$\\($3.49\times10^{3}$)} \\
\bottomrule
\end{tabular}
\caption{Effective parameters ($g_{\mathrm{eff}}$) from CMB constraints, depicting constraints obtained from $P_{\rm ann}$ (top values) and $\sigma_{\rm drag}$  (bottom, in parentheses). $g_{\text{eff}}$ is in units of $(\rm GeV)^{-2}$ for fermionic DM and in $(\rm GeV)^{-1}$ for scalar DM.}\label{tab:tab5}
\end{table}

The upper limits on $P_{\rm ann}$ have been presented in Table~[\ref{tab:constraints}]. 
The relevant expressions for scattering and annihilation cross-sections, which depend on 
the respective lagrangian parameters $g_{\rm eff}$ and the mass of DM, are given in Table~[\ref{tab:effective-operators}]. As mentioned, we are interested in velocity-independent operators 
(i.e, operators $F_5, F_8, F_9, S_1$ given in Table~[\ref{tab:effective-operators}]). 
 
Considering one such operator at a time, constraints on $g_{\rm eff}$ obtained from $P_{\rm ann}$ and $\sigma_{\rm drag}$ are shown in Fig.~\ref{fig:g_eff}. The constraints obtained from 
$P_{\rm ann}$ appears to be more stringent as compared to that obtained from $\sigma_{\rm drag}$. The respective upper bounds on 
various interaction strengths have been depicted in Table~\ref{tab:tab5} 
We observe that the upper limits on the vector operator ($F_5$) and tensor operator ($F_9$) 
decrease with the mass of DM, while the pseudo-vector operator ($F_8$) and scalar DM operator ($S_1$) 
increase with the mass of DM. This behavior can be explained as follows. For fermionic DM  with $m_{\psi} \gtrsim 1$ MeV, the upper bound on effective coupling is proportional to $1/m_{\psi} $ for the vector ($F_5$) and tensor ($F_9$) operators, whereas for the pseudo-vector ($F_8$) operator it is proportional to $ m_{\psi} $.
For scalar DM   ($S_1$) operators the same is proportional $ m_{\phi} $. The detailed expressions have been provided in 
Appendix~\ref{sec:appendixB}.

In the following, we compare the constraints on the effective interaction strengths with similar constraints from direct detection experiments \cite{XENON10:2011prx,DarkSide:2018bpj,XENON:2019gfn,SENSEI:2020dpa,XENON:2021qze,PandaX-II:2021nsg,DarkSide:2022knj}. The relevant constraints are shown in Fig.~\ref{fig:g_eff} along with the direct detection constraints coming from different experiments.  Among the direct detection bounds, the leading constraints from \texttt{SENSEI} and \texttt{DAMIC-M} \cite{SENSEI:2019ibb,SENSEI:2023zdf,DAMIC-M:2023gxo} cover the range from 500 keV to 20 MeV. In the range of 20 MeV to 60 MeV, the strongest bounds come from the \texttt{PANDAX-II} \cite{PandaX-II:2021nsg} experiment, while for DM masses above 30 MeV up to the GeV scale, the leading constraints are provided by the \texttt{XENON-1T} \cite{XENON:2021qze} experiment. The constraint on $g_{\rm eff}$ from direct detection experiments for the mass range 1 MeV – 100 MeV is between $\mathcal{O}(1)$ and $\mathcal{O}(10^{-4})$, respectively, for all given operators. In this mass range, the strongest constraints for all operators come from \texttt{CMB}. Further, the constraints from \texttt{PLANCK} are more stringent than those obtained from indirect searches in the \cite{Cirelli:2023tnx}. 

\section{Conclusions}
\label{sec:conclusion}
In this work, the effect of DM-electron interaction has been considered in the light of 
CMBR data. It has been assumed that such interactions, at the level of 
lagrangian, can be described by effective operators. This description holds good 
when the mediators of such interactions are very heavy compared to the energy 
scale of the relevant scattering or annihilation processes. The presence of 
DM-electron interaction terms lead to annihilation of a pair of DM into a pair of $\bar{e}
e$, as well as scattering of DM particles with free electrons present around the epoch 
of recombination. These processes may be governed by one or more effective 
operators, which may be present in the interaction lagrangian.

Around the recombination epoch, the annihilation of DM can inject energy into the 
plasma, which can enhance the ionization 
of the neutral hydrogen atoms, affecting the optical depth around the recombination 
epoch. Further, such annihilation processes increase the free electron fraction, which 
can affect the polarization of the CMBR photons. DM-electron scattering leads to a non-vanishing drag between these two species. This depletes the matter power at rather 
small length scales. The effect can be opposite at rather larger length scales, which
is possibly due to an enhanced contribution to the baryon loading, as DM is dragged by 
baryons. The effects of DM annihilation and drag have been separately considered in the 
literature, as we discussed. In this analysis, it has been emphasised that, as both of 
these processes stem from the same effective lagrangian terms, therefore, generally, both 
annihilation and scattering need to be considered while considering an interacting DM 
scenario. Note that, for $m_{\chi} \gtrsim 0.51$ MeV, DM annihilation into $\bar{e}-e$ is 
kinematically viable. Thus, an extension of the standard model of cosmology $\Lambda$CDM 
has been studied including two additional parameters $P_{\rm ann}$ and $\sigma_{\rm drag}$ 
representing the effects of DM annihilation and drag with electrons, respectively. 
Only $s$-wave annihilation and velocity-independent drag have been considered in the 
present study. Further, following the literature, DM particles are assumed to have 
an effective temperature $T_{\chi}$ around the epoch of recombination. As DM is very 
weakly coupled to the SM plasma, the respective sound speeds have been assumed 
to be independent. Note that kinetic equilibrium in the DM sector at a temperature 
$T_{\chi}$, generally requires sizable DM-DM interactions around the same epoch. 
While such an interaction may be achieved with very light mediators in the DM sector, 
such additional interactions have not been considered in the present context. We plan 
to address it in a future study. 

We used (a modified version of) the publicly available code \texttt{CLASS} and MCMC code \texttt{MontePython} to estimate the posterior distribution of the relevant 
parameters using the \texttt{Planck}-2018 data set (high-l TTTEEE lite, low-l TT, 
low-l EE, lensing) dataset for the present analysis. We find that the presence of both 
annihilation and drag affects the (95\% c.l.) upper limits on the respective 
parameters $P_{\rm ann}$ and $\sigma_{\rm drag}$. It has been observed that 
the upper limits on $\sigma_{\rm drag}$ is relaxed by $\mathcal{O}(10)\%$, as compared to the scenario with $P_{\rm ann}$ is set to zero. The constraints on $P_{\rm ann}$ shows moderate dependence on mass of DM, especially for light DM with mass of $\mathcal{O}(1)$ MeV - $\mathcal{O}(10)$ GeV, which we considered. A relaxed upper limit on $P_{\rm ann}$ can somewhat reduce the lower limit on $m_{\chi}$ to be a viable thermal DM, assuming that s-wave DM annihilation into $\bar{e}e$ is the only annihilation channel present. 

The posteriors are consistent with the non-interacting DM scenario. The upper limits on $P_{\rm ann}$, as obtained in the presence of $\sigma_{\rm drag}$, have been used to infer upper limits on the strengths of the effective operators describing the relevant DM-electron interactions. The constraints on effective interactions are 
stringent; in particular, for $ 2m_e \lesssim m_{\chi} \lesssim \mathcal{O}(10)$ MeV these dominate over the constraints on the same lagrangian parameters, as inferred from the direct (and indirect) detection experiments.

\section*{Acknowledgements}
RD thanks CSIR, India for financial support through Senior Research Fellowship (File no. 09/ 1128 (13346)/ 2022 EMR-I) and Shiv Nadar IoE (Deemed to be University). AP thanks the Indo-French Centre for the Promotion of Advanced Research for supporting the postdoctoral fellowship through the proposal 6704-4 under the Collaborative Scientific Research Programme. The authors gratefully acknowledge Kimberly Boddy and Vera Gluscevic for initial discussions regarding the code.

\appendix
\section{The Collisional Boltzmann Equation}
\label{sec:appendixA}
We consider the situation where the only relevant process for DM is its 2-to-2 scattering with electrons and annihilation of dark matter to an electron-positron pair. The evolution of the distribution functions is determined by the Boltzmann equation \cite{Cyr-Racine:2015ihg}.
\begin{equation}
\frac{d f_\chi}{d \lambda}= C_{\chi e \leftrightarrow \chi e}[p] + C_{\chi \tilde{\chi} \leftrightarrow \bar{e}e}[p]
\end{equation}
where $f_\chi$  is the distribution function for the evolution of dark matter. 

$$
\frac{\partial f_\chi}{\partial \tau} + \frac{p}{b} \hat{p}^i \frac{\partial f_\chi}{\partial x^i} + p \frac{\partial f_\chi}{\partial p} \left[ -H + \frac{\partial \phi}{\partial \tau} - \frac{E}{p} \hat{p}_i \frac{\partial \psi}{\partial x^i} \right] = \frac{a}{E_\chi} (1 + \psi)\left( C_{\chi e \leftrightarrow \chi e}[p] + C_{\chi \tilde{\chi} \leftrightarrow \bar{e}e} [p]\right) 
$$
We need an equation for its bulk velocity. We multiply both sides by $v_\chi$ and integrate over all $v_\chi$. In non relativistic limits $E_\chi \simeq m_\chi$
\begin{equation}
\begin{aligned}
\begin{split}
&\int \vec{v}_\chi \frac{d^3 v_\chi}{(2 \pi)^3} \left[\frac{\partial f_\chi}{\partial \tau}+\frac{p}{E} \hat{p}^i \frac{\partial f_\chi}{\partial x^i}+p \frac{\partial f_\chi}{\partial p}\left(-\mathcal{H}+\frac{\partial \phi}{\partial \tau}-\frac{E}{p} \hat{p}^i \frac{\partial \psi}{\partial x^i}\right)\right] \\
&= \int \vec{v}_\chi \frac{d^3 v_\chi}{(2 \pi)^3} \left(\frac{a}{m_\chi}(1+\psi) \left( C_{\chi e \leftrightarrow \chi e}[p] + C_{\chi \tilde{\chi} \leftrightarrow \bar{e} e} [p]\right) \right)
\end{split}
\end{aligned}
\label{eq:bolt}
\end{equation}
Solving L.H.S gives
\begin{equation}
    \frac{\partial}{\partial \tau} \left( n^{(0)}_{\chi} \, {v}_{\chi} \right) 
+ \nabla \left[ \frac{\delta p_{\chi}}{m_{\chi}} \right] 
+ 4 \mathcal{H} n^{(0)}_{\chi} {v}_{\chi} 
\end{equation}

$\delta p_{\chi} = c_{\chi}^2 \delta \rho_{\chi} = c_{\chi}^2 n^{(0)}_{\chi} m_{\chi} \delta_{\chi},
$, where $c_{\chi}^2$ is the DM sound speed squared.\\
Using particle conservation and following ref.~\cite{Cyr-Racine:2015ihg} we get left side to be:
\begin{equation}\label{eq:collision}
    \frac{\partial \vec{V}_\chi}{\partial \tau} + c_\chi^2  \vec{\nabla} \delta \chi + \frac{\dot{a}}{a} \vec{V}_\chi
\end{equation}    

where $V_\chi$ is the peculiar velocity of the DM.\\
Both annihilation and scattering are governed by the same operators, but since scattering (or drag) dominates the effect in perturbation, and the effect of annihilation mainly comes from changes in the free electron fraction $x_e$, we can ignore annihilation for now. While annihilation can impact perturbations in some cases, it is not within the scope of this work.
\subsection{Momentum transfer for DM-electron scattering}
In the C.O.M frame (For a single collision)
\[
\begin{aligned}
& \vec{v}_{\chi_{,c}}=\vec{v}_{\chi}-\vec{v}_{\text{cm}} \\
& \vec{v}_{\text{cm}}=\frac{m_{\chi} \vec{v}_{\chi}+m_{b} \vec{v}_{b}}{m_{\chi}+m_{b}} \\
& \vec{v}_{\chi_{,c}}=\vec{v}_{\chi}-\left(\frac{m_{\chi} \vec{v}_{\chi}+m_{b} \vec{v}_{b}}{m_{\chi}+m_{b}}\right) \\
& \vec{v}_{\chi_{,c}}=\frac{m_{b} \vec{v}_{\chi}-m_{b} \vec{v}_{b}}{m_{\chi}+m_{b}} \\
& \vec{v}_{\chi_{,c}}=\frac{m_{b}(\vec{v}_{\chi}-\vec{v}_{b})}{m_{\chi}+m_{b}}
\end{aligned}
\]
Velocity Exchange if after collision dark matter moves in $\hat{n}$ direction C.O.M frame

\[
\begin{aligned}
& \Delta \vec{v}_{\chi}=\vec{v}_{\chi ,c}^{\prime}-\vec{v}_{\chi_{,c}} \\
& \Delta \vec{v}_{\chi}=\frac{m_{b}\left|\vec{v}_{\chi}-\vec{v}_{b}\right|}{m_{\chi}+m_{b}} \hat{n}-\frac{m_{b}(\vec{v}_{\chi}-\vec{v}_{b})}{m_{\chi}+m_{b}}
\end{aligned}
\]
Here $\vec{v}_{\chi_{,c}}$ and $\vec{v}_{\chi,c}^{\prime}$ refers to dark matter velocity after scattering before and after scattering in C.O.M frame.
\[
\begin{aligned}
\Delta \vec{v}_{\chi}=\frac{m_{b}}{m_{\chi}+m_{b}}\left|\vec{v}_{\chi}-\vec{v}_{b}\right|\left(\hat{n}-\frac{(\vec{v}_{\chi}-\vec{v}_{b})}{\left|\vec{v}_{\chi}-\vec{v}_{b}\right|}\right)
\end{aligned}
\]

In non-relativistic limit, $\Delta \vec{P}_{\chi}=m_{\chi} \Delta \vec{v}_{\chi}$
\begin{equation}\label{eq:momentum_exchange}
\Delta \vec{P}_{\chi}=\frac{m_{\chi} m_{b}}{m_{\chi}+m_{b}}\left|\vec{v}_{\chi}-\vec{v}_{b}\right|\left(\hat{n}-\frac{(\vec{v}_{\chi}-\vec{v}_{b})}{\left|\vec{v}_{\chi}-\vec{v}_{b}\right|}\right)
\end{equation}

\subsection{Drag from DM-Electron Scattering}\label{subsubsec:Drag}
The collision term in the Boltzmann equation for DM with initial velocity $v_\chi$ scattering with a baryon with initial velocity $v_b$, resulting in the final velocities $v'_\chi$ and $v'_b$, respectively.
The collision term is given by:

\begin{equation*}
\begin{aligned}
 C_{\chi e \leftrightarrow \chi e}[p] = &\frac{1}{m_\chi} \iiint |M|^{2} \times (2 \pi)^{4} \delta^{4}(P_{\chi} + P_{b} - P_{\chi}^{\prime} - P_{b}^{\prime}) \frac{d^{3} v_{b}}{(2 \pi)^3} \frac{d^{3} v_{\chi}^{\prime}}{(2 \pi)^3} \frac{d^{3} v_{b}^{\prime}}{(2 \pi)^3} \\
& \times \bigg( f_{b}(v_{b}^{\prime}) f_{\chi}(v_{\chi}^{\prime}) [1 - f_{\chi}(v_{\chi})][1 - f_{b}(v_{b})] - f_{\chi}(v_{\chi}) f_{b}(v_{b}) [1 - f_{\chi}(v_{\chi}^{\prime})][1 - f_{b}(v_{b}^{\prime})] \bigg)
\end{aligned}
\end{equation*}

using the value of the collision term in eq.~\ref{eq:bolt}

\begin{equation}
\begin{aligned}
\int \vec{v}_\chi C_{\chi e \leftrightarrow \chi e}[p]\frac{d^{3} v_{\chi}}{(2 \pi)^3} = &\frac{1}{m_\chi} \iiiint |M|^{2} \times (2 \pi)^{4} \delta^{4}(P_{\chi} + P_{b} - P_{\chi}^{\prime} - P_{b}^{\prime}) \vec{v_\chi}\frac{d^{3} v_{\chi}}{(2 \pi)^3}\frac{d^{3} v_{b}}{(2 \pi)^3} \frac{d^{3} v_{\chi}^{\prime}}{(2 \pi)^3} \frac{d^{3} v_{b}^{\prime}}{(2 \pi)^3} \\
& \times \bigg( f_{b}(v_{b}^{\prime}) f_{\chi}(v_{\chi}^{\prime}) [1 - f_{\chi}(v_{\chi})][1 - f_{b}(v_{b})] - f_{\chi}(v_{\chi}) f_{b}(v_{b}) [1 - f_{\chi}(v_{\chi}^{\prime})][1 - f_{b}(v_{b}^{\prime})] \bigg)
\end{aligned}
\label{eq:1}
\end{equation}

From here onwards we will use, \( f_{b}^{\prime} = f_{b}(v_{b}^{\prime}) \), \( f_{\chi}^{\prime} = f_{\chi}(v_{\chi}^{\prime}) \), \( f_{\chi} = f_{\chi}(v_{\chi}) \), and \( f_{b} = f_{b}(v_{b}) \).
For fermionic dark matter and baryons, the collision integral simplifies to:

\begin{equation}
\begin{aligned}
\int \vec{v}_\chi C_{\chi e \leftrightarrow \chi e}[p]\frac{d^{3} v_{\chi}}{(2 \pi)^3} = & \frac{1}{m_\chi}\iiiint |M|^{2} \times \delta^{4}(P_{\chi} + P_{b} - P_{\chi}^{\prime} - P_{b}^{\prime}) \vec{v}_\chi \frac{d^{3} v_{\chi}}{(2 \pi)^3}\frac{d^{3} v_{b}}{(2 \pi)^3} \frac{d^{3} v_{\chi}^{\prime}}{(2 \pi)^3} \frac{d^{3} v_{b}^{\prime}}{(2 \pi)^3} \\
& \times \bigg( f_{b}^{\prime} f_{\chi}^{\prime} - f_{\chi} f_{b} \bigg)
\end{aligned}
\label{eq:2}
\end{equation}

The collision integral can be expressed in terms of the differential cross-section \( \frac{d\sigma}{d\Omega} \):
For elastic scattering $\left| \vec{v}_{\chi} - \vec{v}_{b} \right|$ = $\left| \vec{v}_{\chi}^\prime - \vec{v}_{b}^\prime \right|$
\begin{equation*}
\begin{aligned}
\int \vec{v}_\chi C_{\chi e \leftrightarrow \chi e}[p]\frac{d^{3} v_{\chi}}{(2 \pi)^3} = & \frac{1}{m_\chi}\int \vec{v}_\chi\frac{d^{3} v'_{\chi}}{(2 \pi)^3} \int \frac{d^{3} v'_{b}}{(2 \pi)^3} \int d\Omega \left( \frac{d\sigma}{d\Omega} \right) \left| \vec{v}_{\chi} - \vec{v}_{b} \right| f_{\chi}^{\prime} f_{b}^{\prime} \\
& -\frac{1}{m_\chi} \int \frac{d^{3} v'_{\chi}}{(2 \pi)^3} \int \frac{d^{3} v'_{b}}{(2 \pi)^3} \int d\Omega \left( \frac{d\sigma}{d\Omega} \right) \left| \vec{v}_{\chi} - \vec{v}_{b} \right| f_{\chi} f_{b}
\end{aligned}
\end{equation*}

\begin{equation}
\begin{aligned}
\int \vec{v}_\chi C_{\chi e \leftrightarrow \chi e}[p] \frac{d^{3} v_{\chi}}{(2 \pi)^3} = & \frac{1}{m_\chi}\int (\vec{v}_{\chi}^{\prime} - \Delta \vec{v}_{\chi}) \frac{d^{3} v'_{\chi}}{(2 \pi)^3} \int \frac{d^{3} v'_{b}}{(2 \pi)^3} \int d\Omega \left( \frac{d\sigma}{d\Omega} \right) \left| \vec{v}^\prime_{\chi} - \vec{v}^\prime_{b} \right| f_{\chi}^{\prime} f_{b}^{\prime} \\
& - \frac{1}{m_\chi}\int \vec{v}_{\chi} \frac{d^{3} v'_{\chi}}{(2 \pi)^3} \int \frac{d^{3} v'_{b}}{(2 \pi)^3} \int d\Omega \left( \frac{d\sigma}{d\Omega} \right) \left| \vec{v}^\prime_{\chi} - \vec{v}^\prime_{b} \right| f_{\chi} f_{b}
\end{aligned}
\label{eq:4}
\end{equation}

since $v_\chi^\prime,v_b^\prime$ are dummy indices in the first term. so replacing $v_\chi^\prime \to v_\chi$ and $v_b^\prime \to v_b$ and using conservation of phase space $\int \frac{d^{3} v_{\chi}}{(2 \pi)^3} \int  \frac{d^{3} v_{b}}{(2 \pi)^3} = \int \frac{d^{3} v'_{\chi}}{(2 \pi)^3} \int  \frac{d^{3} v'_{b}}{(2 \pi)^3}$

\begin{equation*}
\begin{aligned}
\int \vec{v}_\chi C_{\chi e \leftrightarrow \chi e}[p] \frac{d^{3} v_{\chi}}{(2 \pi)^3} = & \frac{1}{m_\chi}\int (\vec{v}_{\chi} + \Delta \vec{v}_{\chi}) \frac{d^{3} v_{\chi}}{(2 \pi)^3} \int \frac{d^{3} v_{b}}{(2 \pi)^3} \int d\Omega \left( \frac{d\sigma}{d\Omega} \right) \left| \vec{v}_{\chi} - \vec{v}_{b} \right| f_{\chi} f_{b} \\
& - \frac{1}{m_\chi}\int \vec{v}_{\chi} \frac{d^{3} v_{\chi}}{(2 \pi)^3} \int \frac{d^{3} v_{b}}{(2 \pi)^3} \int d\Omega \left( \frac{d\sigma}{d\Omega} \right) \left| \vec{v}_{\chi} - \vec{v}_{b} \right| f_{\chi} f_{b}
\end{aligned}
\label{eq:4b}
\end{equation*}

This simplifies to:

\begin{equation}
\int \vec{v}_\chi C_{\chi e \leftrightarrow \chi e}[p] \frac{d^{3} v_{\chi}}{(2 \pi)^3} = \frac{1}{m_\chi}\int \Delta \vec{v}_{\chi} \frac{d^{3} v_{\chi}}{(2 \pi)^3} \int \frac{d^{3} v_{b}}{(2 \pi)^3} \int d\Omega \left( \frac{d\sigma}{d\Omega} \right) \left| \vec{v}_{\chi} - \vec{v}_{b} \right|
\label{eq:5}
\end{equation}

If there are \( n_{b} \) scattering centers, the momentum transfer becomes:

\begin{equation*}
\begin{aligned}
\int \vec{v}_\chi C_{\chi e \leftrightarrow \chi e}[p] \frac{d^{3} v_{\chi}}{(2 \pi)^3} = & \frac{n_{b}}{m_{\chi}} \int \frac{d^{3} v_{\chi}}{(2 \pi)^3} f_{\chi} \int \frac{d^{3} v_{b}}{(2 \pi)^3} f_{b}\int d\Omega \left( \frac{d\sigma}{d\Omega} \right) \left| \vec{v}_{\chi} - \vec{v}_{b} \right| \Delta \vec{P}_{\chi}
\end{aligned}
\label{eq:6}
\end{equation*}

Substituting the momentum transfer cross-section and simplifying gives:

\begin{equation}
\begin{aligned}
 \int \vec{v}_\chi C_{\chi e \leftrightarrow \chi e}[p] \frac{d^{3} v_{\chi}}{(2 \pi)^3} =& \frac{\rho_b}{m_\chi + m_b}\int \frac{d^{3} v_{\chi}}{(2 \pi)^3} f_{\chi} \int \frac{d^{3} v_{b}}{(2 \pi)^3} f_{b} \int d\Omega \left( \frac{d\sigma}{d\Omega} \right) \left| \vec{v}_{\chi} - \vec{v}_{b} \right|^{2} \\
& \times \left( \hat{n} - \frac{(\vec{v}_{\chi} - \vec{v}_{b})}{\left| \vec{v}_{\chi} - \vec{v}_{b} \right|} \right)
\end{aligned}
\end{equation}

The distribution functions for dark matter (\( f_{\chi} \)) and baryons (\( f_{b} \)) are given by:
\[
f_{\chi}(\vec{v}_{\chi}) = \frac{1}{(2\pi)^{3}\bar{v}_{\chi}^{3}} \exp\left[-\frac{(\vec{v}_{\chi} - \vec{V}_{\chi})^{2}}{2\bar{v}_{\chi}^{2}}\right],
f_{b}(\vec{v}_{b}) = \frac{1}{(2\pi)^{3}\bar{v}_{b}^{3}} \exp\left[-\frac{(\vec{v}_{b} - \vec{V}_{b})^{2}}{2\bar{v}_{b}^{2}}\right],
\]
where \( \bar{v}_{\chi}^{2} = T_{\chi}/m_{\chi} \) and \( \bar{v}_{b}^{2} = T_{b}/m_{b} \) are the thermal velocity dispersions.

To simplify calculations, we introduce new variables:
\[
\vec{v}_{m} \equiv \frac{\bar{v}_{b}^{2} \vec{v}_{\chi} + \bar{v}_{\chi}^{2} \vec{v}_{b}}{\bar{v}_{b}^{2} + \bar{v}_{\chi}^{2}}, 
\vec{v}_{r} \equiv \vec{v}_{\chi} - \vec{v}_{b}.
\]

With these new variables, the distribution functions remain factorizable:
\[
\int d^{3}v_{\chi} \, f_{\chi}(\vec{v}_{\chi}) \int d^{3}v_{b} \, f_{b}(\vec{v}_{b}) = \int d^{3}v_{r} \, f_{r}(\vec{v}_{r}) \int d^{3}v_{m} \, f_{m}(\vec{v}_{m}).
\]

The new distribution functions \( f_{m} \) and \( f_{r} \) are Gaussian:
\[
f_{m}(\vec{v}_{m}) = \frac{1}{(2\pi)^{3}\bar{v}_{m}^{3}} \exp\left[-\frac{(\vec{v}_{m} - \vec{V}_{m})^{2}}{2\bar{v}_{m}^{2}}\right],\hspace{0.1cm}
f_{r}(\vec{v}_{r}) = \frac{1}{(2\pi)^{3}\bar{v}_{r}^{3}} \exp\left[-\frac{(\vec{v}_{r} - \vec{V}_{r})^{2}}{2\bar{v}_{r}^{2}}\right],
\]

where:
\[
\vec{V}_{m} = \frac{\bar{v}_{b}^{2} \vec{V}_{\chi} + \bar{v}_{\chi}^{2} \vec{V}_{b}}{\bar{v}_{b}^{2} + \bar{v}_{\chi}^{2}},\hspace{0.5cm}
\bar{v}_{m}^{2} = \frac{\bar{v}_{\chi}^{2} \bar{v}_{b}^{2}}{\bar{v}_{\chi}^{2} + \bar{v}_{b}^{2}}
\]
\[
\vec{V}_{r} = \vec{V}_{\chi} - \vec{V}_{b}, \hspace{0.5cm}
\bar{v}_{r}^{2} = \bar{v}_{\chi}^{2} + \bar{v}_{b}^{2}.
\]
$V_\chi$ and $V_b$ are the peculiar velocities of the DM and the baryon. In this work, we are interested in the interaction of a DM-electron, so we now replace b with e for the electron.
\begin{equation}
    \int \vec{v}_{\chi} C_{\chi e \leftrightarrow \chi e}[p] \frac{d^{3} v_{\chi}}{(2 \pi)^3} =-\frac{Y_b \rho_e \sigma_{\rm drag}}{m_\chi+m_e} \int \frac{d^3 v_r}{(2 \pi)^3} f_r\left(\vec{v}_r\right) v_r^{n+1} \vec{v}_r \int \frac{d^3 v_m}{(2 \pi)^3} v_m f_m\left(\vec{v}_m\right) 
\end{equation}

where we obtain the second line by completing the integration over angles to obtain the momentum-transfer cross section, and by utilizing. The integral over $\vec{v}_m$ simply evaluates to 1 ,  and the remaining integral over $\vec{v}_r$ yields the result
$$
\int \vec{v}_{\chi} C_{\chi e \leftrightarrow \chi e}[p] \frac{d^{3} v_{\chi}}{(2 \pi)^3} =-\frac{Y_b \rho_e \sigma_{\rm drag} \mathcal{N}_n}{m_\chi+m_e} \bar{v}_r^{n+1} \vec{v}_{r},{} _1F_1\left(-\frac{n+1}{2}, \frac{5}{2},-\frac{r^2}{2}\right) .
$$
using Eq.~[\ref{eq:collision}]
\begin{equation}\label{eq:velocity_equation}
    \frac{\partial \vec{V}_\chi}{\partial \tau} + c_\chi^2  \vec{\nabla} \delta \chi + \frac{\dot{a}}{a} \vec{V}_\chi = 
-a  Y_b \rho_e \sigma_{\rm drag} \frac{\mathcal{N}_n}{m_\chi + m_e} 
\bar{v_r}^{n+1}  \vec{v}_r, {}_1F_1 
\left(
-  \frac{n+1}{2}, \frac{5}{2}, -\frac{r^2}{2}
\right)
\end{equation}

Taking divergence $\nabla\cdot$, both sides then in k space above equation can
be written as
\[
\dot{\theta}_{\chi} +\mathcal{H}\theta_{\chi} - c^2_{\chi} k^2 \delta_{\chi} = R_{\chi} (\theta_b - \theta_{\chi})
\]

where 
\begin{equation}\label{eq:momentum-transfer}
R_\chi = a  Y_b \rho_e \sigma_{\rm drag} \frac{\mathcal{N}_n}{m_\chi + m_e} 
\bar{v_r}^{n+1} {}_1F_1 
\left(
-  \frac{n+1}{2}, \frac{5}{2}, -\frac{r^2}{2}
\right)
\
\end{equation}

Here, the relative velocity is given by  
\begin{equation}  
\bar{v_r} = \left( \frac{T_b}{m_e} + \frac{T_{\chi}}{m_{\chi}} + \frac{V^2_{\rm rms}}{3} \right)^{1/2}.  
\label{eq:bulkv}
\end{equation}  

where $V^2_{\rm rms} = \langle \vec{V}^2_{\chi} \rangle_{\xi} =
\int \frac{dk}{k} \, \Delta_{\xi} \left( \frac{\theta_b - \theta_c}{k} \right)^2$ and $\langle \dots \rangle_{\xi}$ denote an average with respect to the primordial curvature perturbation, and  
$\Delta_{\xi} \simeq 2.4 \times 10^{-9}$ is the primordial curvature variance per $\log k$ \cite{PhysRevD.82.083520}.

For $n \geq 0$, the root mean square (r.m.s.) velocity is zero, i.e., $V_{\rm r.m.s} = 0$. Since we are primarily interested in velocity-independent scattering cross-sections in this work, the expression simplifies to  
$ 
\bar{v_r} = \left( \frac{T_b}{m_e} + \frac{T_{\chi}}{m_{\chi}} \right)^{1/2}.  
$

\section{Cross-section for Annihilation and Scattering}
\label{sec:appendixB}
As mentioned in Sec.~\ref{sec:pert2}, we will continue to denote (Dirac) fermionic dark matter by $\psi$ and scalar dark matter by $\phi$ in this section. Assuming both dark matter and electrons are non-relativistic, the differential dark matter-electron scattering cross section is given by:

\[
\frac{d\sigma_{\rm drag}}{d\cos\theta_*} = \frac{\overline{|\mathcal{M}|}_{sc}^2}{32\pi(m_e+m_\psi)^2}
\]

\begin{align} \label{eq:drag}
\sigma_{\rm drag}=\int \frac{\overline{|\mathcal{M}|}_{sc}^2}{32 \pi\left(m_e+m_\psi\right)^2}  d \cos \theta_* \
\end{align}

where $\overline{|\mathcal{M}|}_{sc}^2$ is the spin-averaged amplitude squared of dark matter-baryon scattering and $\sigma_{\rm drag}$ is the cross-section of DM-e scattering.

Similarly, for DM annihilation to $\bar{e}e$ the estimated cross section is given by:

\begin{align}\label{eq:ann}
    \frac{d\sigma_{ann}}{d\cos\theta_*} = \frac{\overline{|\mathcal{M}|}_{\rm ann}^2}{16\pi v_{\psi, rel} s} \sqrt{1 - \frac{{m_e}^2}{{m_\psi}^2}}
\end{align}
where \(\overline{|\mathcal{M}|}_{\rm ann}^2\) is the spin-averaged amplitude squared for dark matter annihilation into electrons. The Mandelstam variable \(s\) is defined in the center-of-mass (c.o.m.) frame as  
\[
s \simeq 4m_\psi^2 + m_\psi^2 v_{\psi, rel}^2
\]
where \(v\) is the relative velocity of the annihilating DM particles.

First, we will try to solve the Axial-Vector operator for Dark Matter Electron interaction.
\begin{align*}
\mathcal{L} = {g}_{eff} \bar{\psi} \gamma^\mu \gamma^5 \psi \bar{e} \gamma_\mu \gamma^5 e \\
\end{align*}

Eqs.~[\ref{eq:drag}] and ~[\ref{eq:ann}] give the cross-section for dark matter scattering and annihilation as:

\begin{align}
\sigma_{\rm drag}=g_{\text {eff }}^2 \frac{3}{4}\left(\frac{16 m_e^2}{\pi\left(1+\frac{m_e}{m_\psi}\right)^2}\right)
\end{align}
and,
\begin{align}\label{equation:7.4}
\langle \sigma v_{\psi, rel} \rangle=\frac{1}{2 \pi} g_{\rm eff}^2 m_\psi^2\left(\sqrt{1-\frac{m_e^2}{m_\psi^2}}\right)\left[\frac{m_e^2}{m_\psi^2}+\frac{1}{12}\left(2-\frac{m_e^2}{m_\psi^2}\right) v_{\psi, rel}^2\right]
\end{align}
Similarly, for the Vector operator 
\begin{align*}
\mathcal{L} = {g}_{\rm eff} \bar{\psi} \gamma^\mu \psi \bar{e} \gamma_\mu e \\
\end{align*} 
Eqs.~[\ref{eq:drag}] and ~[\ref{eq:ann}] give the cross-section for dark matter scattering and annihilation as:
\begin{align}
\sigma_{\rm drag}= \frac{g_{\text{eff}}^2 \mu^2}{\pi }
\end{align}

and,
\begin{equation}
\begin{aligned}
\langle \sigma v_{\psi, rel} \rangle = & g_{\text{eff}}^2 \frac{ m_\psi^2}{2 \pi } \sqrt{1 - \frac{m_e^2}{m_\psi^2}} \times \left[\left(2 + \frac{m_e^2}{m_\psi^2}\right)  \right] \\
\end{aligned}
\end{equation}
where $\mu$ is reduced where $\mu = \frac{m_e m_\psi}{m_e + m_\psi}$ for dark matter and electron system.

For the Tensor operator
\begin{align*}
\mathcal{L} = {g}_{\text{eff}} \bar{\psi} \sigma^{\mu \nu} \psi \bar{e} \sigma_{\mu \nu} e \\
\end{align*} 
again, following the same process, the cross-section for dark matter scattering and annihilation for the above tensor operator as: 

\begin{align}
\sigma_{\rm drag}= \frac{48 \times \mu^2 g_{\rm eff}^2}{\pi }
\end{align}

and,
\begin{equation}
\begin{aligned}
\langle \sigma v_{\psi, rel} \rangle= & g_{\rm eff}^2\frac{ m_\psi^2}{2 \pi }  \sqrt{1 - \frac{m_e^2}{m_\psi^2}} \times \left[ 16 \left(1 + \frac{m_e^2}{m_\psi^2}\right) \right].
\end{aligned}
\end{equation}
where $\mu$ is reduced where $\mu = \frac{m_e m_\psi}{m_e + m_\psi}$ for dark matter and electron system. These equations can also be checked from \cite{Buckley:2011kk,Kumar:2013iva}.

For the scalar operator, the interaction lagrangian is:  

\begin{align*}
    \mathcal{L} = {g}_{\text{eff}} \phi^\dagger \phi \bar{e} e
\end{align*}  

Using Eqs.~[\ref{eq:drag}] and ~[\ref{eq:ann}], the cross-section for dark matter scattering and annihilation is:  

\begin{align}\label{equation:7.9}
\sigma_{\rm drag}= \frac{g_{\text{eff}}^2 }{16 \pi (1 + \frac{m_\phi}{m_e})^2}
\end{align}  

and,  

\begin{equation}
\begin{aligned}
\langle \sigma v_{\phi, rel} \rangle= & g_{\text{eff}}^2 \frac{1}{8 \pi } \left( {1 - \frac{m_e^2}{m_\phi^2}}\right)^{\frac{3}{2}} 
\end{aligned}
\end{equation}  

where $\mu = \frac{m_e m_\phi}{m_e + m_\phi}$ is the reduced mass of the dark matter-electron system. 

\vspace{1cm}

\bibliographystyle{utphys}

\bibliography{reference.bib}

\end{document}